\documentclass[final,authoryear,5p,times,twocolumn]{elsarticle}

\usepackage{graphicx}

\usepackage{amssymb}
\usepackage{amsmath}

\usepackage{enumerate}
\usepackage{times}
\usepackage{jtb}

\DeclareMathAlphabet{\bi}{OML}{cmm}{b}{it}

\journal{Journal of Theoretical Biology}

\begin{document}

\begin{frontmatter}



\title{Species assembly in model ecosystems, II: Results of the assembly
process.}


\author[leg]{Jos\'e A.\ Capit\'an}
\ead{jcapitan@math.uc3m.es}
\author[leg]{Jos\'e A.\ Cuesta\corref{cor}} \ead{cuesta@math.uc3m.es}
\cortext[cor]{Corresponding author.} 
\author[sev]{Jordi Bascompte} \ead{bascompte@ebd.csic.es}
\address[leg]{Grupo Interdisciplinar de Sistemas Complejos (GISC), 
Departamento de Matem\'aticas, Escuela Polit\'ecnica Superior, 
Universidad Carlos III de Madrid, E-28911 Legan\'es, Madrid, Spain}
\address[sev]{Integrative Ecology Group, Estaci\'on Biol\'ogica de 
Do\~nana, Consejo Superior de Investigaciones Cient\'{\i}ficas,
c/ Americo Vespucio s/n, E-41092 Sevilla, Spain}

\begin{abstract}

In the companion paper of this set \citep{capitan:2010a} we have 
developed a full analytical treatment of the model of species
assembly introduced in \cite{capitan:2009}. This model is based on 
the construction of an assembly graph containing all viable configurations
of the community, and the definition of a Markov chain whose transitions 
are the transformations of communities
by new species invasions.  In the present paper we provide 
an exhaustive numerical analysis of the model, describing the average 
time to the recurrent state, the statistics of avalanches, and the 
dependence of the results on the amount of available resource.  Our 
results are based on the fact that the Markov chain provides an asymptotic 
probability distribution for the recurrent
states, which can be used to obtain averages of observables as well as 
the time variation of these magnitudes during succession, in an exact
manner. Since the absorption times into the recurrent set are found 
to be comparable to the
size of the system, the end state is quickly reached (in units of the 
invasion time). Thus, the final ecosystem can be regarded as a fluctuating complex 
system where species are continually replaced by newcomers without ever 
leaving the set of recurrent patterns. The assembly graph is dominated by 
pathways in which most invasions are accepted, triggering small extinction 
avalanches. Through the assembly process, communities become less resilient 
(e.g., have a higher return time to equilibrium) but become more robust 
in terms of resistance against new invasions. 

\end{abstract}


\end{frontmatter}

\section{Introduction}
\label{intro}

Understanding the mechanisms leading to species assembly in ecological 
communities is a challenging issue. In particular, assembly models have 
been used to understand the observation that natural communities are both 
complex and stable \citep{mccann:2000,dunne:2006}. 

Assembly models try to mimic the sequential arriving of rare species (invaders)
to which natural communities are subjected. Standard assembly models 
\citep{drake:1990,law:1993,law:1996} use ``species pools'' as (finite) sets 
of potential invaders. Pools are usually defined by labeling species according 
to some niche variable (usually a species trait like body size) and then drawing
randomly their interactions from predetermined probability distributions \citep{law:1996}.
Sequential invaders of any given resident community are selected from the pool 
at each invasion attempt, and the resulting community after the invasion can be
determined according to some population dynamics.
For models using Lotka-Volterra equations, the permanence \citep{hofbauer:1998}
of the invaded community is a suitable criterion
which determines the same final community as the 
numerical integration of the equations \citep{morton:1996}.

The most remarkable results of previous assembly models are: (i) a final end state 
is eventually reached, which can be either a 
single community or a cycle involving several communities \citep{morton:1997},
(ii) average species richness (complexity) increases with successional time
\citep{post:1983,drake:1990,law:1996}, and
(iii) stability, understood as resistance against invasions, also increases with time
\citep{case:1990,law:1996,morton:1997}. Thus assembly models conform a 
well-founded theoretical framework
that provides a positive relationship between stability and complexity in model
communities.

In a previous paper \citep{capitan:2009} we have 
provided a picture of the assembly process of an ecosystem
as a Markov chain evolving in a
certain configuration space. This space is made of all viable communities
for a given set of parameters (resource saturation, interspecific competition,
consumption rates\dots). The invasion process by a new species induces transitions
as a result of the perturbations created in the community by the newcomer. The 
process drives the community to an end state resistant to invasions.
For some parameter values 
this end state is just a single uninvadable community. For the remaining values,
the end state is formed by a set of 
communities with equal number of trophic levels and
similar number of species per level, which transform
into each other as a result of new invasions. In this set, communities can
always be invaded but they never abandon the set. These complex
end states are a generalization of the end cycles found in previous assembly
models \citep{morton:1997}, and the fact that they had not been observed so
far is probably due to 
limitations in the pool of invaders of these previous models. In the preceding
paper \citep{capitan:2010a}
we have shown that the existence of these complex end states is a result
of a top predator attempting to invade a community when its establishment is
not allowed by the parameters of the model.

Our model recovers the main findings of previous assembly models 
\citep{post:1983,drake:1990,case:1990,law:1993,law:1996},
such as the resistance of end states against invasions, or the increase of
complexity (biodiversity) along the assembly \citep{capitan:2009}.
Its main virtue is therefore being sufficiently simple so as to allow
mapping out all assembly pathways, thus providing a global picture of the
assembly process. 

Despite recovering the above similar results, there are important differences 
between our work and early  
models. First, the niche variable in our model is simply the trophic level 
\citep{capitan:2009},
which renders our species pool infinite (in contrast to most previous models;
but see \cite{post:1983} for an exception).
However, interactions in the pool are averaged over each
trophic level under a species symmetry assumption \citep{capitan:2010a},
which decreases substantially the number of different assembly
pathways. Second, in our model the permanence of the final community is
guaranteed because we are able to show that equilibrium communities
are globally stable \citep{hofbauer:1998} under the assumption of neutrality
within each trophic level. And third, in standard assembly models
magnitudes are averaged over a set
of stochastic realizations of the process of sequential invasions,
where invaders are randomly chosen from the species in the pool
not yet present in the community. Since we
are able to map out all the invasion pathways for this model, we do not need
to resort to average magnitudes over realizations but we can calculate them
exactly \citep{capitan:2009}. Even for our simple model, the number of 
possible pathways is too high to be accounted by through simulation 
[see details in \cite{capitan:2009}].
This is one of the main advantages of our model 
with respect to former ones, which in turn allows us to establish its independence
on history. The uniqueness of the end state for these kind of
models was not proven until now, although it was already found that most of 
the simulated assembly sequences led to a single set of final communities 
\citep{morton:1997}.

In the first paper of this suite \citep{capitan:2010a} we have performed a 
detailed analysis of the analytical properties of the Lotka-Volterra
population dynamics underlying our assembly
model, as well as the stability properties
of the interior equilibria from the dynamic point of view.
Our communities represent a mean-field version of trophic networks: the feeding 
relations are assumed to take place only between
contiguous trophic levels and the strength of each interaction is averaged
to a uniform value. This assumption of symmetry allowed
us to simplify the differential equations, showing that
in our model the set of species numbers at each level
$\{s_{\ell}\}_{\ell=1}^L$ is enough to determine the equilibrium densities 
and the dynamics of a community with $L$ trophic levels. 

The present paper presents new results of the assembly process ranging from structural 
properties of the
building of stable communities to dynamical properties of the stochastic invasion
process. The transition matrix associated to the chain 
allows us to define a stationary distribution of probabilities over the assembly graph. 
We will use that distribution to calculate averages of biologically relevant observables
in the ecosystem, like the average number of species, the total population density,
etc., and even to obtain distributions of certain magnitudes like the fraction
of extinct species after a destructive invasion. 
But this is not the only advantage of our approach. The transition matrix also provides
the time evolution of the probability, so the dependence in time of any magnitude
can be obtained exactly. Due to its simplifications, the model reduces the number
of possible communities to a finite graph. Once we have that graph, the theory
of finite Markov chains can be exploited to obtain the dynamical properties
of the assembly process.

\begin{figure}[t!]
\begin{center}
\includegraphics[width=80mm,clip=true]{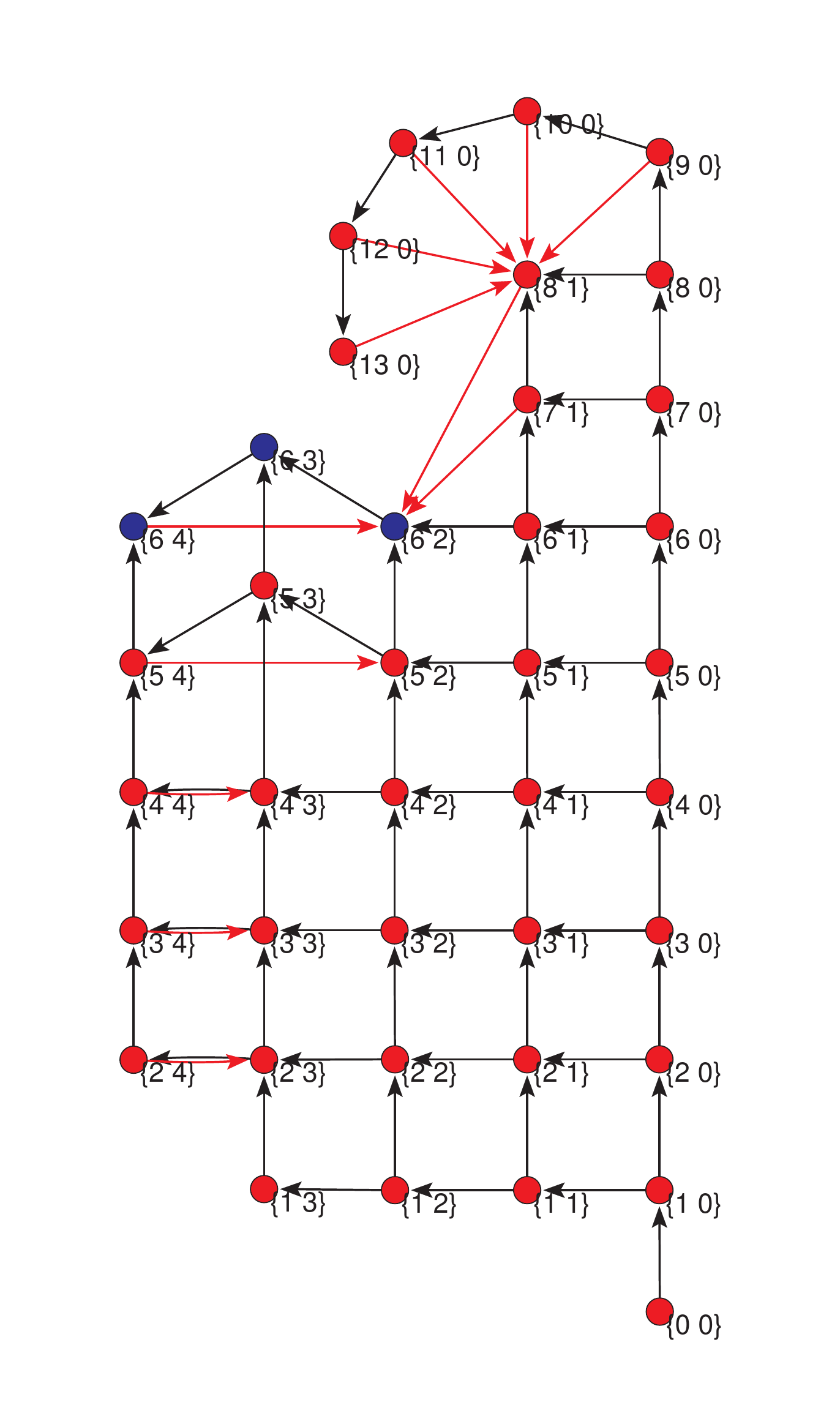}
\caption{(Color online) Assembly graph obtained for a value of the resource
saturation $R=80$. It is made of $39$ communities (nodes), 
each of them with either one or two trophic levels. Transitions shown with a black
arrow indicate that the invasion is accepted, and those with a red arrow refer to
a rearrangement in the resulting community after the invasion. Transient nodes are
filled in red, and recurrent nodes are filled in dark blue. In this case, the final 
end state of recurrent communities comprises 3 communities forming an end cycle like
those found by \cite{morton:1997}. Labels of each node show
the species numbers $\{s_1,s_2\}$ of each trophic level in the community.
(For the choice of parameters see Section~\ref{ss:resource}.)}
\label{fig:chain}
\end{center}
\end{figure}

The paper has been
organized as follows. In Section~\ref{s:markov} we revisit the definition
of the Markov chain associated to the assembly process and show that 
all communities can be classified into transient or recurrent.
Section~\ref{s:results} is devoted to discuss the main statistical
results that can be obtained for our model, for instance, the asymptotic distribution
within the complex end states (Section~\ref{ss:asym}), the dependence of averages upon
variation of the resource saturation (Section~\ref{ss:resource}), 
the dependence of the results upon variation of the parameters
of the model (Section~\ref{ss:param}), the average 
time to reach the end state (Section~\ref{ss:absor}), the statistics
of avalanches of extinctions caused by invasions (Section~\ref{ss:extin}) or
the variation of biologically relevant averages with successional time 
(Section~\ref{ss:ave}). We finally discuss our findings and
their implications in Section~\ref{s:discussion}.

\section{The assembly process as a Markov chain}
\label{s:markov}

Let us start with a detailed description of the Markov chain associated to
our assembly model 
In the first paper of this suite \citep{capitan:2010a} we showed that, solving 
the linear system that defines the interior equilibrium point of the
Lotka-Volterra equations (i.e., the system obtained by equating to zero the
per-capita population growth rates $\dot{n}_i/n_i$, with $i$ running over
the set of all species) we can determine all viable communities 
(i.e., those whose population densities are above a certain extinction
threshold $n_c>0$) that are compatible with a given set of parameters. 
Although in principle
the population model allows for infinitely many species at
each level, it turns out that the set of viable communities
is finite. This is a consequence of the existence of the extinction threshold, 
that precludes the appearance of arbitrarily small populations
(an unrealistic feature of deterministic population dynamics models). 
There is another limitation due to the finite amount of abiotic
resource that maintains our model communities. 
In our previous work we modeled this resource with a linear functional
response with a saturation value $R$ \citep{capitan:2009,capitan:2010a}.
$R$ accounts for the amount of resource that would be reached in the absence
of consumers. On the
one hand there is a maximum number of levels allowed for a given resource
saturation $R$ \citep{capitan:2010a}; on the other hand population densities
at each level decrease as $1/s_{\ell}$, with $s_{\ell}$ the number of species
in that level \citep{capitan:2010a}, so we can
have populations infinitely close to 
zero. Therefore, the existence of the extinction threshold renders the set of 
communities under consideration finite, and then the associated Markov chain
has a finite number of states.
Besides this being a more realistic description of an ecosystem,
it also drastically simplifies the analysis of the assembly process.

\begin{figure}
\begin{center}
\includegraphics[width=93mm,clip=true]{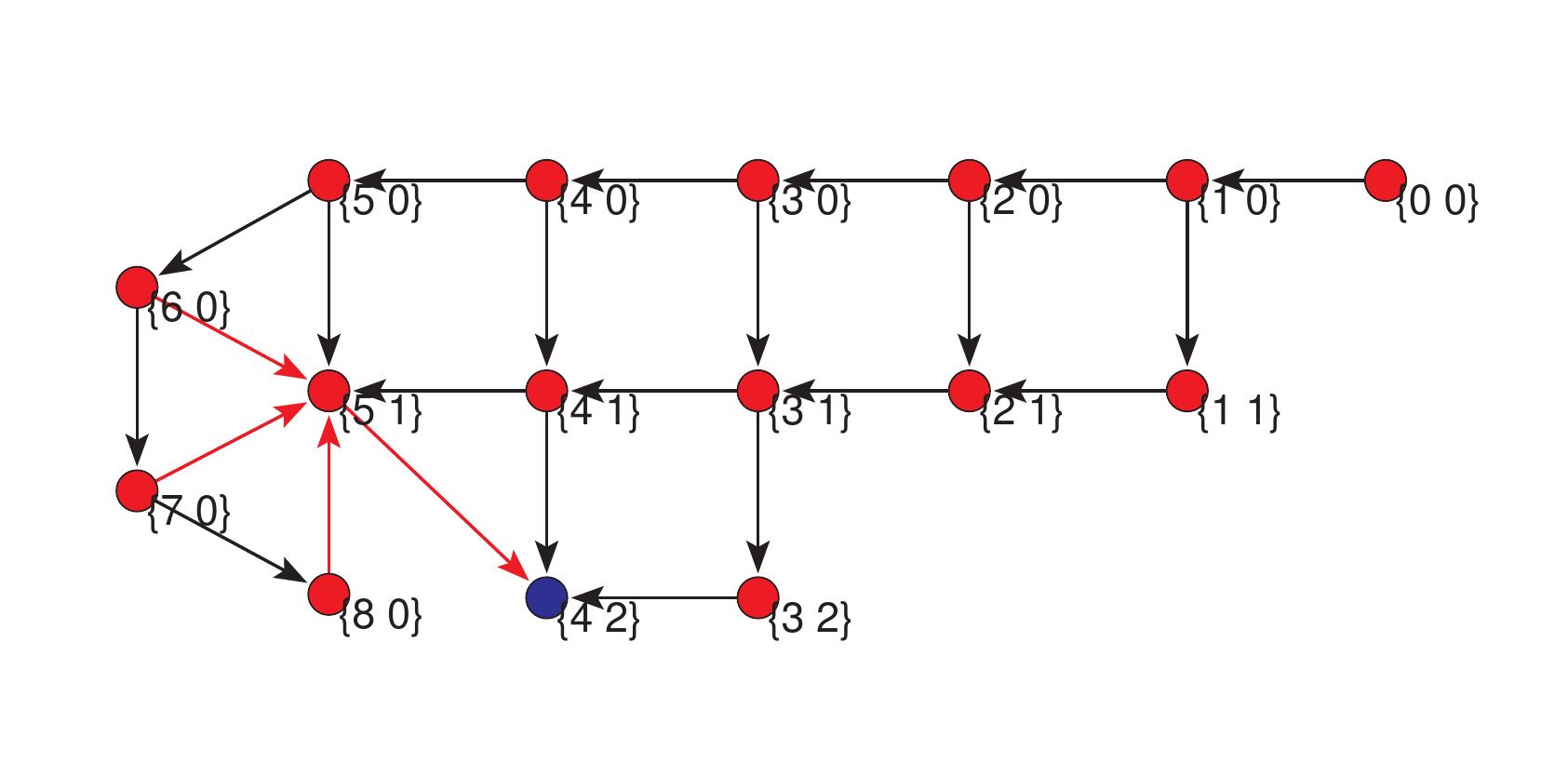}
\caption{(Color online) Same as Figure~\ref{fig:chain} for $R=50$.
The total number of communities in this graph is $16$, and they have up to
$2$ trophic levels. An end state with a single community (with
occupancies $\{4,2\}$) is reached in this case.}
\label{fig:chain1}
\end{center}
\end{figure}

Thus for any choices of parameters there is a \emph{finite} set of viable
communities ---that we denote by $\mathcal{F}$. There will be a link
from community $i$ to community $j$ of the set $\mathcal{F}$ provided
the former is transformed into the latter as a result of an invasion.
Invasions are assumed to occur at a 
uniform rate $\xi$. We assume that the typical dynamical time is much
smaller than $\xi^{-1}$, the mean time between invasions, so that communities
are always at equilibrium when an invasion occurs
[for the validity of this assumption 
see the discussion in \cite{capitan:2010a}]
The population of the invader is assumed as small as possible, i.e.
equal to $n_c$ \citep{roughgarden:1974, turelli:1981}.

\begin{figure*}
\begin{center}
\raisebox{73mm}{(a)}\includegraphics[width=90mm,clip=true]{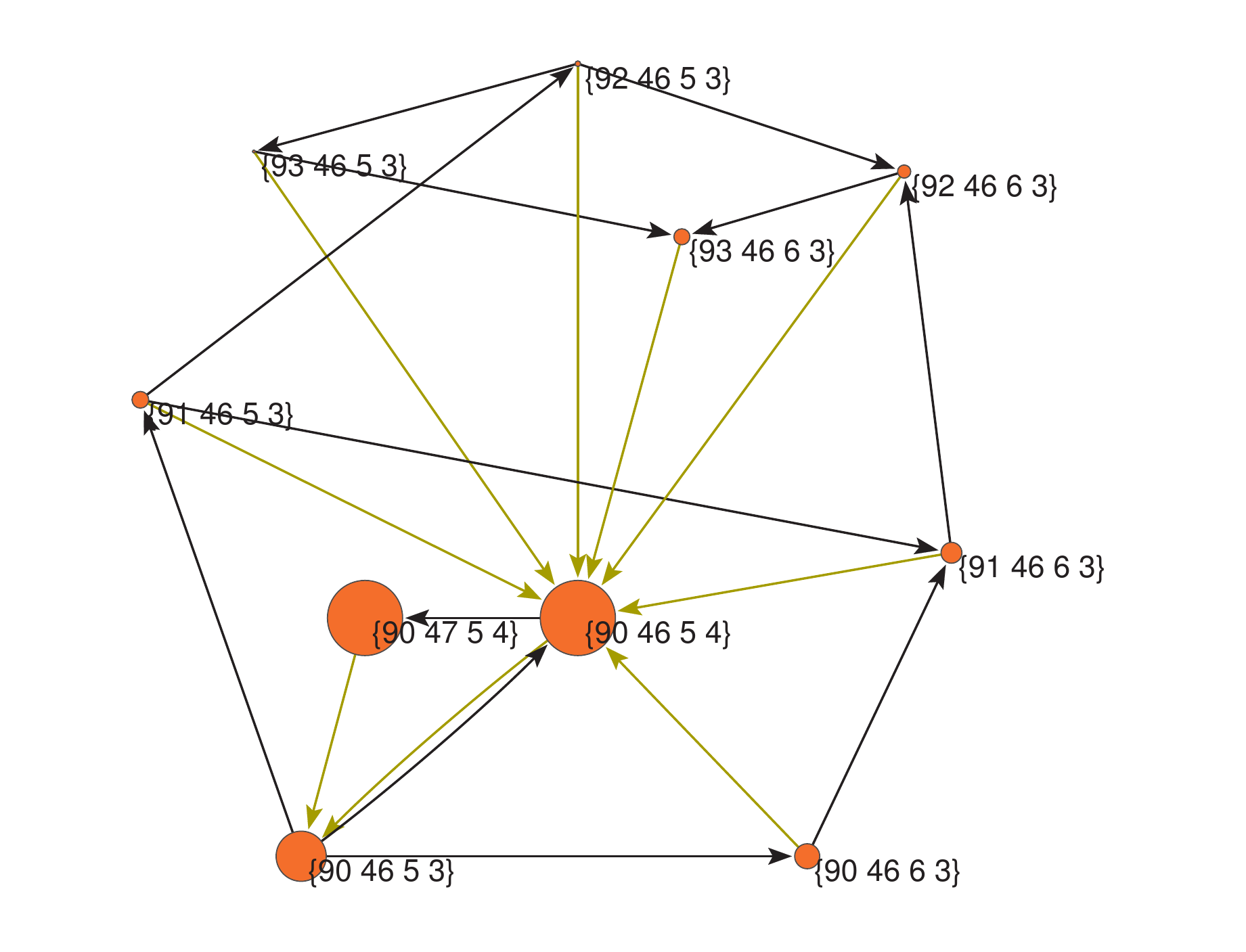}\hspace*{-1mm}
\raisebox{73mm}{(b)}\includegraphics[width=90mm,clip=true]{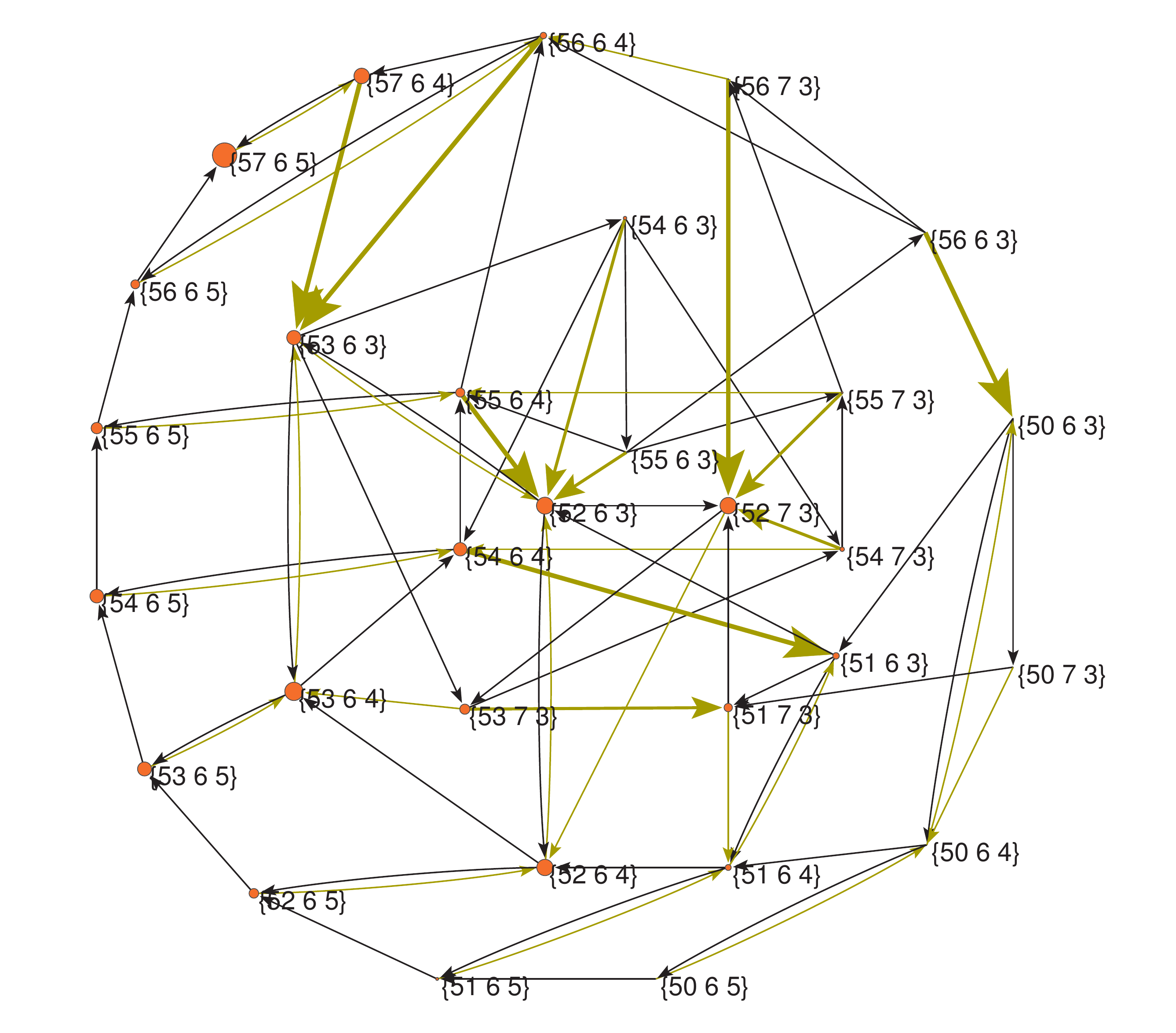}
\caption{(Color online) Graphs of the recurrent subset for values of the
resource saturation $R=990$ (a) and $R=390$ (b)
(other parameters are set as in Figures~\ref{fig:chain} and \ref{fig:chain1}).
The first graph contains 10 communities with 4 trophic levels, whereas the second has 
30 communities with 3 levels. 
For a graphical representation of a recurrent subset more complex than these 
see \cite{capitan:2009}.
All the communities in both cases have three trophic levels. The diameter
of the nodes is proportional to its asymptotic probability. Black arrows show accepted
invasions and yellow ones those causing a reconfiguration (the thickness of each line is
proportional to the relative number of extinct species). Labels indicate the number of
species in each trophic level.}
\label{fig:closed}
\end{center}
\end{figure*}

Consider a community $i\in\mathcal{F}$, with $L$ trophic levels, at its
equilibrium point. Potential invaders are species of level $\ell=1,\dots,L+1$
(species of higher levels would not be able to feed from the existing levels). We randomly
choose $\ell$ and introduce a new species at level $\ell$ of the community $i$.
The extended community will evolve to the interior equilibrium 
corresponding to the new number of species $s_{\ell}+1$ at level $\ell$ \citep{capitan:2010a}. 
If this equilibrium is viable, then the invader is accepted in the resident
community. 
The new community $j$ will also be in $\mathcal{F}$ and a directed link will
go from $i$ to $j$. 
If the new equilibrium is not viable then we apply the procedure discussed
in \cite{capitan:2010a} to determine what is the 
sequence of extinctions until the community becomes viable. Two things can 
happen: either the first level to fall below the extinction threshold is the 
invaded one, or it
is another one. In the former case the invader is simply rejected and 
the community remains unaltered; in the latter, the extinction sequence will lead
to a new community $k\in\mathcal{F}$, and a link will go from $i$ to $k$.

The assembly graph, $\mathcal{G}$, is defined as the connected component
containing the empty community, $\varnothing$, of the directed graph whose nodes
are the elements of $\mathcal{F}$ and whose links are the transitions obtained
by the invasion process just described. Obviously, the way to construct
$\mathcal{G}$ is to start off from $\varnothing$, and proceed by attempting all
possible invasions for every community reached along the assembly process [see 
Figures~\ref{fig:chain} and \ref{fig:chain1} for a graphic representation of 
simple assembly graphs, for a larger value of $R$ see Figure 1 in \cite{capitan:2009}].
The exhaustive characterization of the set of nodes in $\mathcal{G}$ is a bit
demanding. For example, 
the storage of all the communities appearing along the process has been
carried out by using a binary tree \citep{knuth:1997}, by exploiting the mapping
between any configuration $\{s_{\ell}\}_{\ell=1}^L$ and a binary number (we simply
concatenate the binary representations of each species number $s_{\ell}$
to form a branch of the binary tree). Despite this, we have been able to analyze 
graphs with around $10^6$ communities within.

The connection of the species assembly process with a Markov chain on the graph 
$\mathcal{G}$ amounts to assigning certain transition probabilities to each link of 
the assembly graph. We define these probabilities in a simple way. Invaders arrive
at each community at a constant rate $\xi$, independent of the 
level of invasion, and the stochastic process is updated in discrete time (each
time unit is the average time elapsed between consecutive invasions).
Thus, if $i$ and $j$ are two nodes of
$\mathcal{G}$ connected by a link, we assign it the transition probability
\begin{equation}\label{eq:mat}
P_{ij} = \delta_{ij} + \xi Q_{ij},
\end{equation}
where $\delta_{ij} = 1$ if $i=j$ and $0$ otherwise. The matrix elements
$Q_{ij}$ are given by
\begin{equation}
Q_{ij} = \frac{n_{ij}}{L+1}, \quad i\ne j, \qquad
Q_{ii} = -\sum_{j\ne i} Q_{ij},
\end{equation}
where $n_{ij}$ is the number of different invasions of $i$ that lead to $j$
and $L+1$ is the number of different invasions of $i$, provided it has $L$ trophic
levels. Therefore, the probability of the transition $i \rightarrow j$ 
between different communities is 
proportional to the relative frequency of the transition among all the possible
transitions starting from $i$, the invasion rate being the proportionality
constant. The diagonal of $Q$ is chosen so that $P = (P_{ij})$ is a stochastic
matrix.

Since the diagonal elements of the transition matrix $P $ are non-zero, the Markov
chain can not be periodic \citep{karlin:1975}. As the set of viable ecosystems 
$\mathcal{F}$ is finite, $P$ defines the transition matrix of a finite,
aperiodic, Markov chain. The states of one such chain are either transient or
recurrent \citep{karlin:1975}.
There can be one or several subsets of recurrent states, the chain
being ergodic in each of them. Every recurrent subset is a different
end state of the assembly process. The end state of an ecosystem 
will be history-dependent only if there
are at least two such recurrent subsets. Ergodicity implies
that there is a stationary probability distribution on the states of these
subsets which determines the frequency with which the process visits each
of them. [For a full account on Markov chains see e.g.\ \cite{karlin:1975}].
Our model only exhibits a unique recurrent set for any given
set of parameter values \citep{capitan:2009}. 

\begin{table*}
\begin{center}
\begin{tabular}{ccl}
\hline\hline
$parameter$ & $value$ & $interpretation$\\
\hline
$R$ & $10\le R \le 1700$ & Saturation value, in the absence of predation, of the abiotic 
resource abundance\\
$\alpha$ & $1$ & Average mortality rate of consumers\\
$\gamma_-$ & $5$ & Average rate of decrease in preys population caused 
by their being predated\\
$\gamma_+$ & $0.5$ & Average rate of increase in predators population due to feeding\\
$\rho$ & $0 \le \rho \le 1$ & Relative magnitude between intra-- and interspecific
competition\\
$n_c$ & $1$ & Extinction threshold\\
\hline\hline
\end{tabular}
\caption{Summary of parameters of the model and ecological meaning
of each one of them.}
\label{tab:param}
\end{center}
\end{table*}

\begin{figure}
\begin{center}
\includegraphics[width=80mm,clip=true]{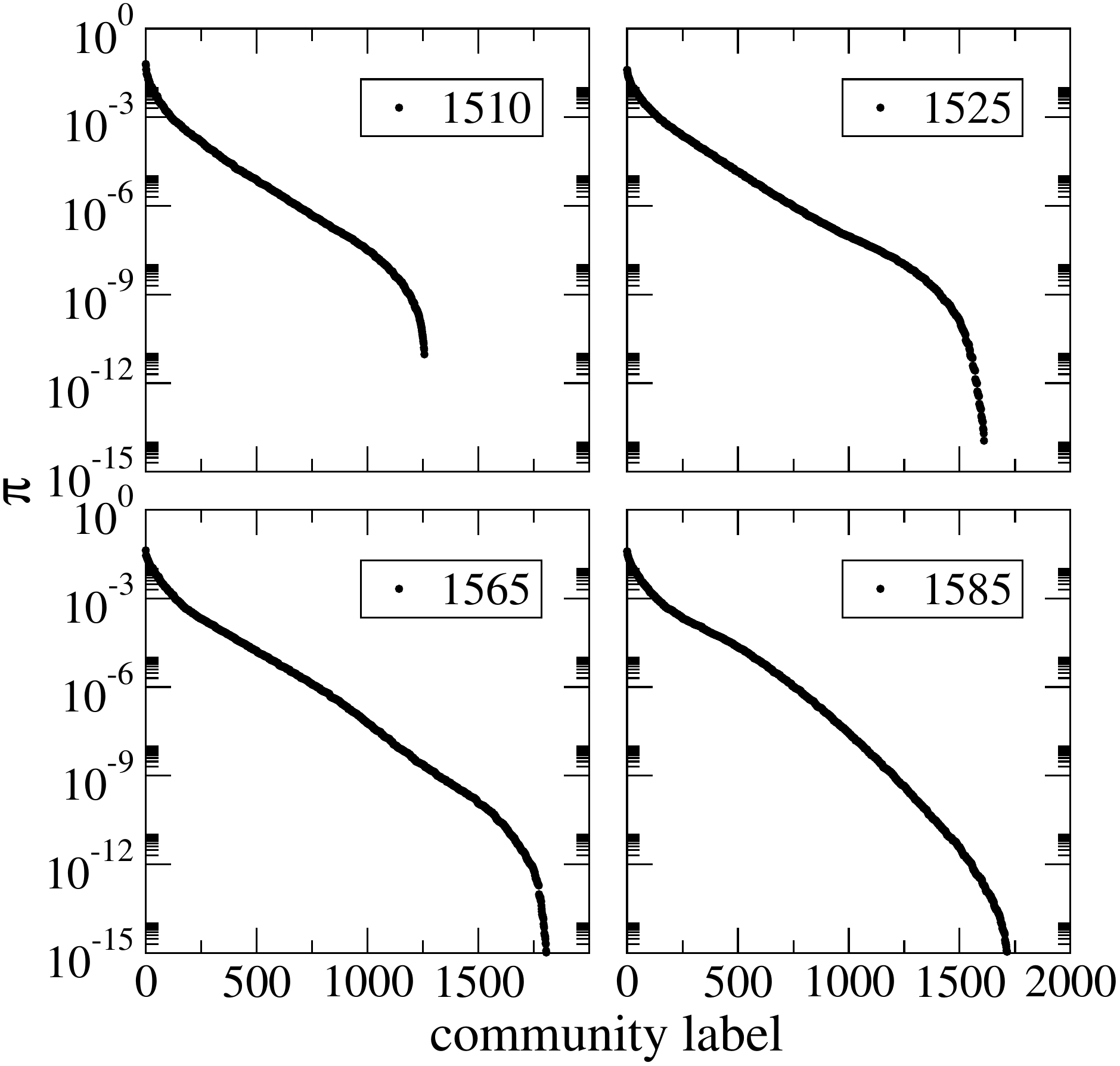}
\caption{Distribution of asymptotic probabilities $\pi$ 
within each recurrent set,
for several values of $R$. Communities are labelled in decreasing order
of probability. The distributions look exponential with a cutoff.}
\label{fig:equi}
\end{center}
\end{figure}

This concludes the definition of the Markov model for the assembly process. 
As we are able to compute the whole transition matrix $P$, 
we have a \emph{complete} and \emph{exact} characterization of
the assembly process. In particular, by selecting an initial state for the
Markov chain (in our case the process always starts off from $\varnothing$), 
we can obtain the evolution of any magnitude 
---numerically but exactly--- without resorting to taking averages over
realizations of the process. In the following section we will discuss
in detail the results that can be obtained from the analysis of the Markov
chain [a brief account of which were reported in \cite{capitan:2009}].

\section{Results}
\label{s:results}

\subsection{Asymptotic distribution}
\label{ss:asym}

To separate transient and recurrent states,
we have applied an algorithm provided by \cite{xie:1998}.
Notice that the characterization of transient and recurrent states
in a finite chain depends only on the graph, not on the transition
probabilities. Only one subset of
recurrent states was found for each set of parameters. 
Let $\mathcal{R}$ denote the
subgraph of $\mathcal{G}$ formed by this ergodic set.
Figure~\ref{fig:closed} shows two examples of these subgraphs.
The particular transition probabilities assigned to 
each link would determine the asymptotic probability distribution
within the recurrent set, but not the subset of nodes contained in it.

In order to calculate the asymptotic probability $\pi_i$ for a community
$i \in \mathcal{G}$, we need to solve the linear system $\pi = \pi P$
\citep{karlin:1975}, in
other words, the vector $\pi$ of probabilities is the left eigenvector 
of the matrix $P$ with eigenvalue 1. 
Since our graphs are very sparse, standard numerical
techniques for solving sparse systems have been applied. 
The eigenvector is normalized to satisfy the condition
$\sum_{i\in\mathcal{R}} \pi_i = 1$. Obviously, we only need to solve
this system for the subgraph corresponding to the recurrent set, since by
definition the asymptotic probability $\pi_i=0$ for any transient state $i$.
Note that our matrix of transition probabilities \eqref{eq:mat} reduces
the condition to be satisfied by $\pi$ to $\pi Q = 0$,
i.e., $\pi$ is a left eigenvector of $Q$  with eigenvalue $0$. It is 
worth noticing that neither the asymptotic distribution, nor the recurrent
subset depends on the invasion rate.

\begin{figure}[t!]
\begin{center}
\includegraphics[width=88mm,clip=true]{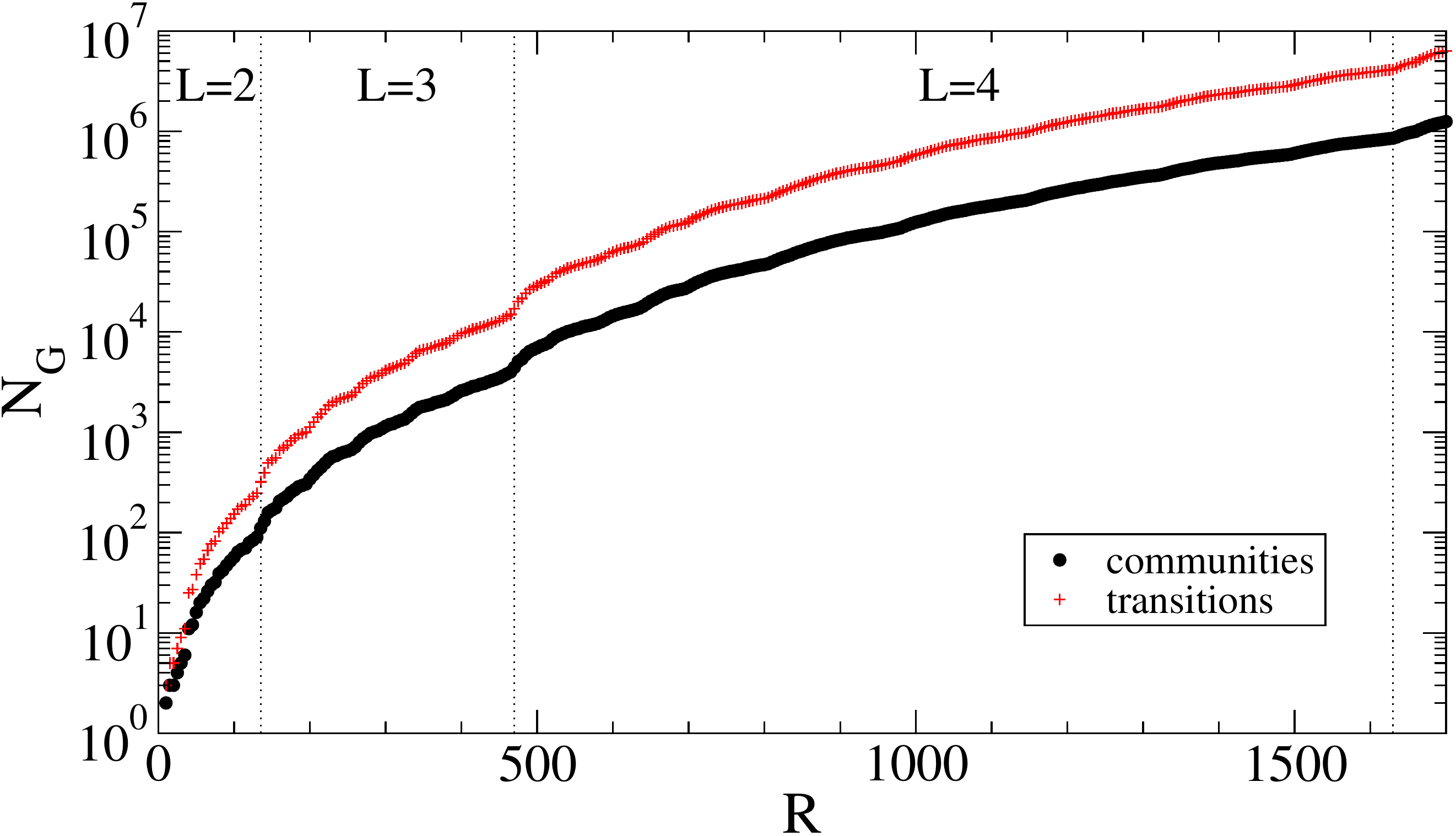}
\caption{(Color online) Total number of communities (black circles, below)
and transitions (red crosses, above)
in the Markov chain as a function of the resource saturation $R$.}
\label{fig:states}
\end{center}
\end{figure}

We can thus obtain a probability distribution for each recurrent
set. Having this probability distribution is therefore
equivalent to defining a statistical mechanics
over the set of viable communities, if we regard $\mathcal{F}$ as the 
phase space of our system. In Figure \ref{fig:equi} we have 
plotted the histogram of probabilities for several values of the resource
saturation $R$ (for these values the number of communities in each set is
larger than $10^3$). Communities are labelled in decreasing order of
probability. These distributions are found to be roughly exponential
over several orders of magnitude, this meaning that only a small
number of communities (in general very similar to each other)
occur with high probability.
These are the communities in which it is more likely to find the ecosystem.
Nonetheless ergodicity implies 
that all communities in the end state are visited with non-zero probability. 
The ecosystem is thus in a complex state, with fluctuating species numbers
in each level due to some invasions being accepted and some others 
causing avalanches of extinctions.

This distribution can be used to calculate the asymptotic average 
over $\mathcal{R}$ of any relevant magnitude $M_i$ defined for every
community, like for instance the average number of
species, the total population, etc. We just need to evaluate
$\langle M\rangle_{\mathcal{R}} = \sum_{i\in \mathcal{R}} \pi_i M_i$.

\subsection{Dependence with the resource saturation}
\label{ss:resource}

\begin{figure}[t!]
\begin{center}
\includegraphics[width=88mm,clip=true]{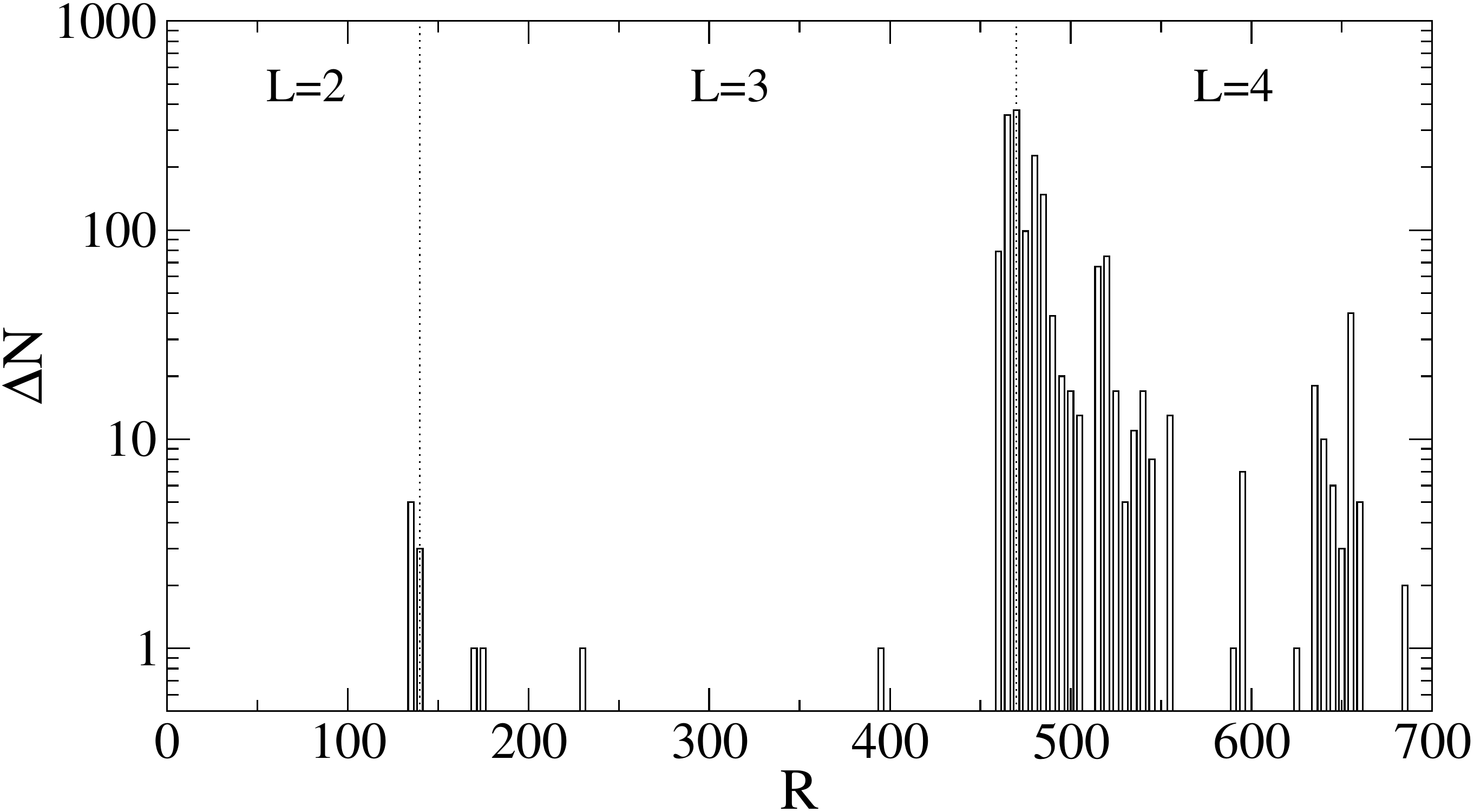}
\caption{Number of viable states that are not reachable trough invasions starting from
the empty community $\varnothing$, $\Delta N = N_{\mathcal{F}}-N_{\mathcal{G}}$. 
Typically there is an accumulation of these communities in the regions
where $R$ allows a top predator in them.}
\label{fig:out}
\end{center}
\end{figure}

All results presented here have been obtained with a death rate $\alpha=1$,
an extinction threshold $n_c=1$, an average rate of increase in 
predators population per predation event $\gamma_+=0.5$, 
and an average decrease in
reproduction rate of preys per predation event $\gamma_-=5$ (see
\cite{capitan:2010a} for details on the way these parameters enter the
population equations of the model 
and Table \ref{tab:param} for a brief summary of them). The
assumption of $\gamma_+ \ll \gamma_-$ is ecologically sound, because
many preys must be consumed to produce a new predator, while loosing
one prey requires a single predation event. A common choice for
the energy transfer between trophic levels is about 10\% \citep{pimm:1991}.
We have checked that the model is robust against variations of these parameters
within reasonable bounds.

In most cases we have taken the ratio of direct inter- to intraspecific competition
$\rho=0.3$. However, we have explored the effect of this parameter in detail
in Section~\ref{ss:param}. 

We have obtained all assembly graphs in a range 
of resource saturations that
goes from $R=10$ up to $R=1700$ with increments $\Delta R=5$. No viable community
is found below $R=10$. The number of communities $N_{\mathcal{G}}$ 
in these graphs goes from just
one (for $R=10$) up to about $10^6$. We have found empirically that both this number 
and the total number of transitions in each graph grow
roughly as $e^{\kappa\sqrt{R}}$, see Figure \ref{fig:states}. 
The maximum number of trophic levels that are allowed up to $R=1700$ is $5$.

We have checked whether the set of communities in the assembly graph is the
whole set $\mathcal{F}$. Given the estimation of the resource values that allow
a maximum number of levels $L_{\rm max}$ [see \cite{capitan:2010a}], 
we have checked the viability of all possible combinations of species numbers
$\{s_{\ell}\}_{\ell =1}^{L_{\rm max}}$ with $L_{\rm max}+1$ 
levels up to a total number of species $S_{\rm max}$ equal to twice the
maximum number of species allowed for that $R$ value. Since there is a huge number of
these combinations when $R$ increases, we have checked this up to $R=700$.
Figure \ref{fig:out} shows the difference $\Delta N=N_{\mathcal{F}}-N_{\mathcal{G}}$. 
In nearly all cases the set of communities
in the assembly graph is $\mathcal{F}$, but we have found several instances ---all
of them near the values of $R$ at which new levels arise---
in which $\mathcal{F}$ contains communities not reachable through the assembly
process, just like in the experiment of \cite{warren:2003}. 
The largest 
difference is found at $R=470$, where $N_{\mathcal{G}}=4800$ and 
$\Delta N=375$, so the highest relative difference reaches 8\%.

\begin{figure}[t!]
\begin{center}
\includegraphics[width=83mm,clip=true]{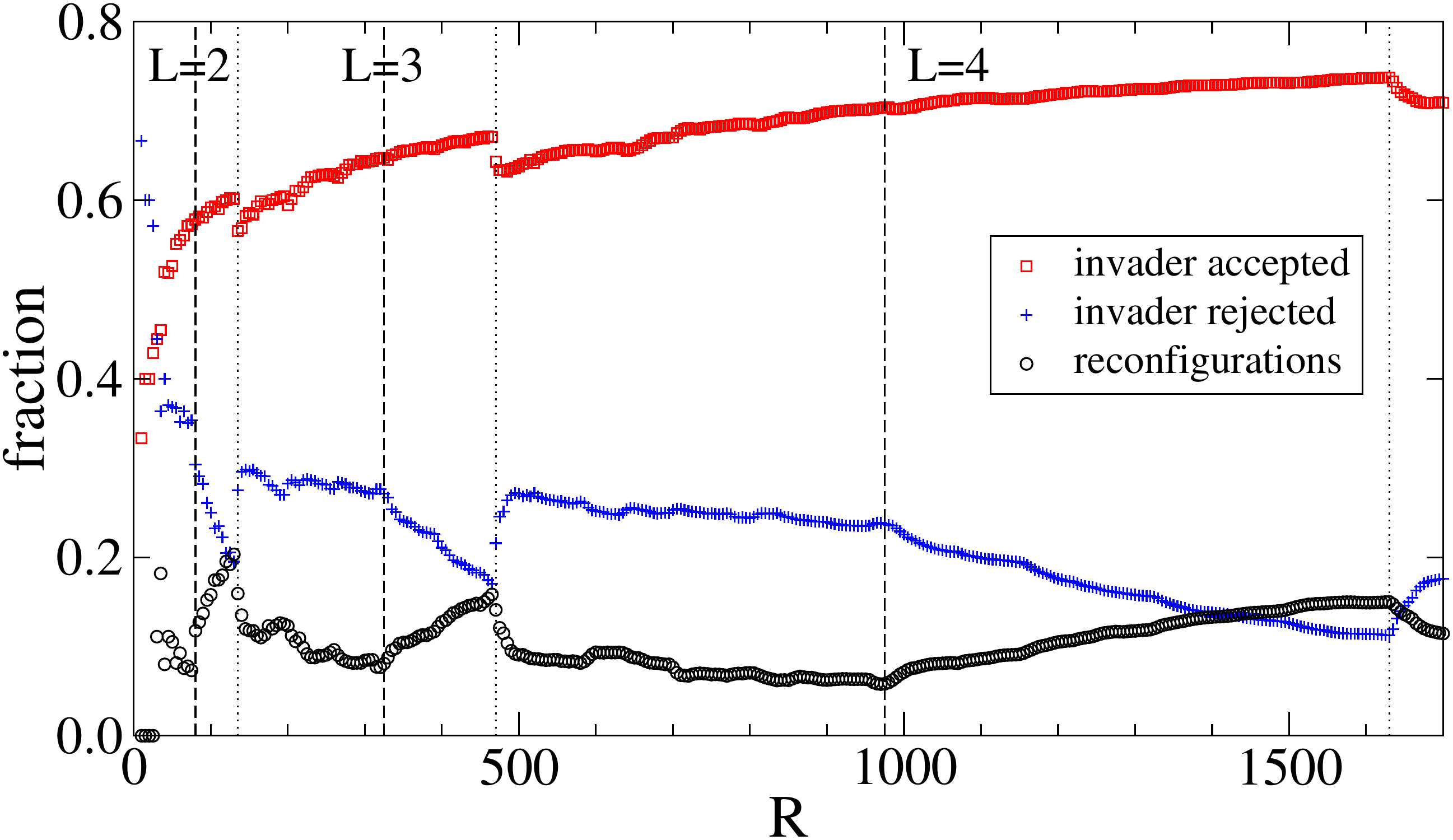}
\caption{(Color online) Statistics of links corresponding
to accepted invasions (red squares), rejected invasions (blue crosses) and
invasions that lead to a reconfigured community through an avalanche of
extinctions (black circles). Dotted lines
correspond to the onsets of acceptance of a new trophic level, and dashed lines
to the beginning of the regions of complex end states.}
\label{fig:recurrent}
\end{center}
\end{figure}

For each $R$ we determined the number of recurrent states of the chain
(see Figure~3 in \cite{capitan:2009} for a plot of this number as a function
of $R$).
We always found a single connected graph, which implies that the end state of
the assembly process does not depend on history for this model \citep{drake:1990}.
This is consistent with the results of \cite{morton:1997} as well
as the experiments of \cite{warren:2003}. 
There are values of $R$ for which this set consists of a unique absorbing state
(or just a few, sometimes forming a cycle), but when $R$ is reaching the values
at which a new trophic level appears, the size of this set increases considerably
(the largest set found contains around $1800$ communities; a tiny fraction of
the whole assembly graph, anyway).
After crossing these values the size of the recurrent set drops again down to just
one absorbing state. \cite{morton:1997} also obtained complex end states
in $6$ out of the $80$ pools they explored, with a number of communities ranging
from $6$ to $138$. 

\begin{figure}
\begin{center}
\vspace*{-1mm}\raisebox{46mm}{(a)}\includegraphics[width=84mm,clip=true]{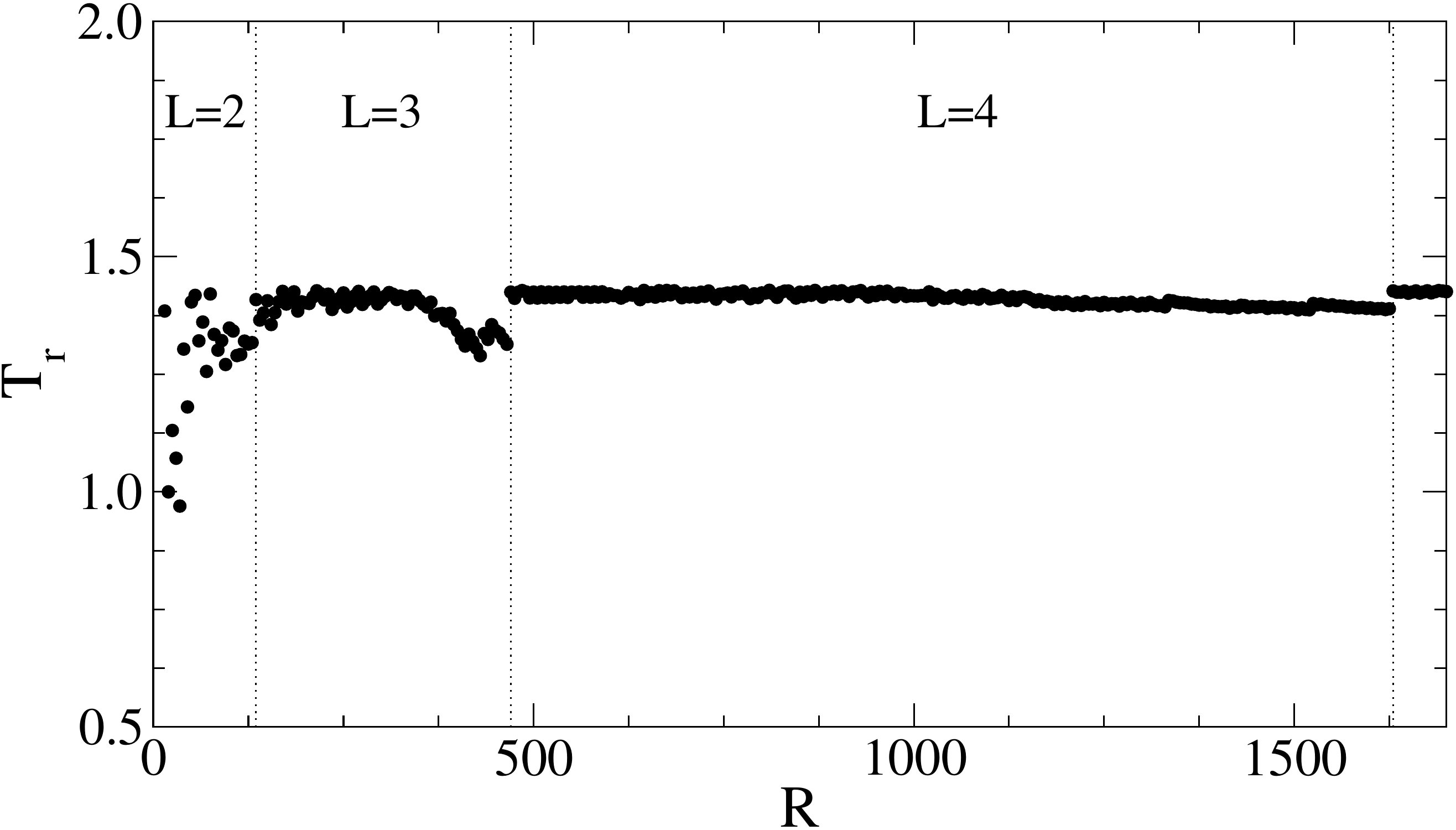}\vspace*{1.5mm}
\raisebox{37mm}{(b)}\hspace*{1.5mm}\includegraphics[width=82mm,clip=true]{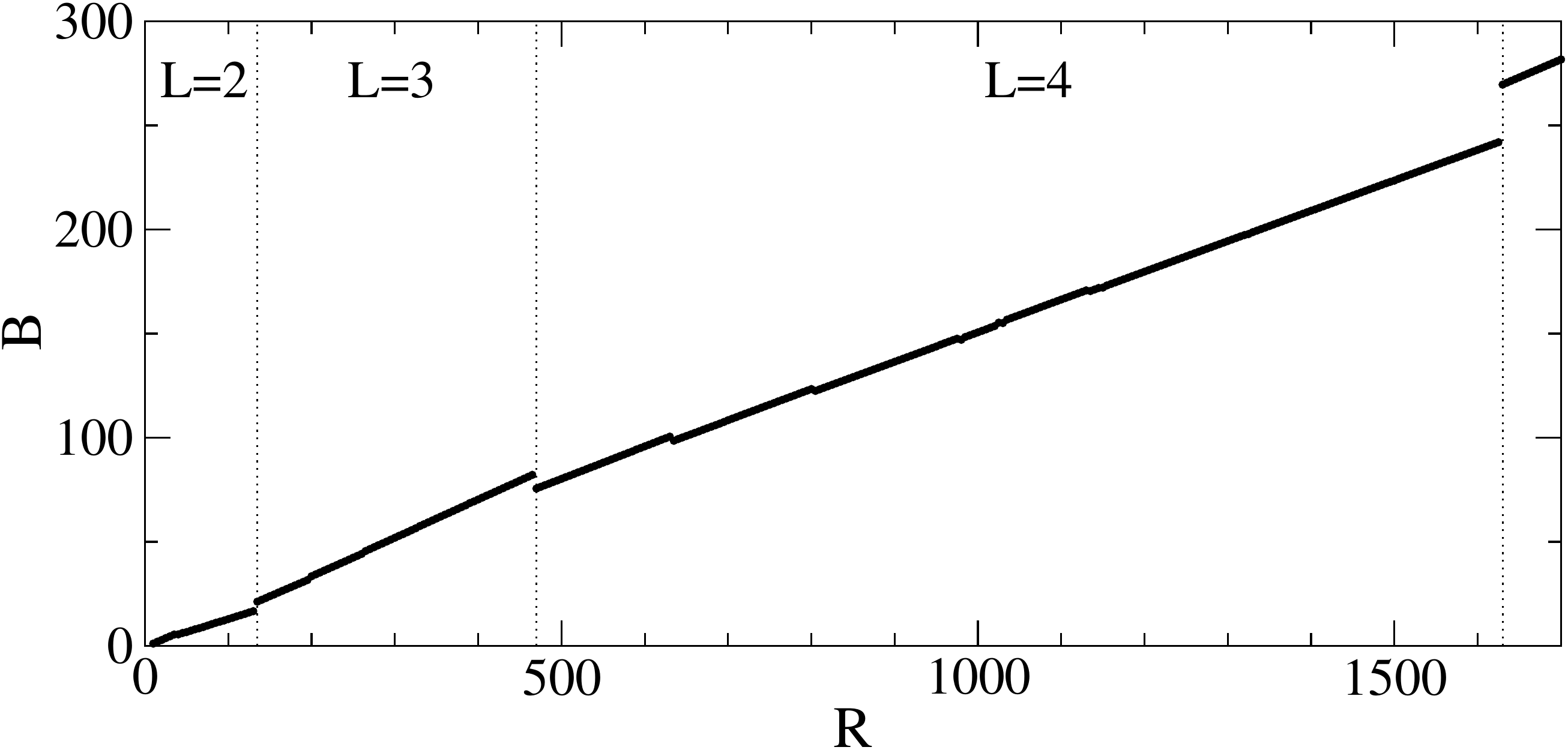}
\caption{(a) Mean return time in the stationary state vs.\ resource saturation
$R$. The behavior is approximately constant, except for the region of low
resources, where the graphs contain less communities and there is more variability.
(b) Mean population density of the community vs.\ $R$.}
\label{fig:return}
\end{center}
\end{figure}

In Figure~\ref{fig:recurrent} we show the fractions of links in the assembly
graph corresponding to invasions that are accepted, rejected, or cause a
reconfiguration in the system through a sequence of extinctions.
The most frequent case is the acceptance of the invader, although there are
around 20\% of rejections and reconfigurations. We can see an increasing trend
to reconfigurations when $R$ corresponds to a complex end state [see 
Figure~3 in \cite{capitan:2009}]. The invasibility criterion
discussed in \cite{capitan:2010a} explains why we observe
an increasing number of rearranged communities in these regions.

As for the dynamic stability (resilience), we can measure the return time,  
i.e., the mean time that a perturbed ecosystem needs to restore equilibrium
\citep{pimm:1977}, averaged over the probability distribution of
the stationary state. It can be obtained
as $T_{\rm r}=-\lambda_{\rm max}^{-1}$, where $\lambda_{\rm max}$ 
is the largest real part of the eigenvalues
of the linear stability matrix ---which is always negative in our communities
since they are globally stable. We observe that this time 
is roughly independent on the end state,
being approximately constant as a function of the resource saturation
$R$ (see Figure~\ref{fig:return}a).

For each end state, regardless on whether it is an absorbing community or a recurrent set, we have 
calculated another average. In Figure~\ref{fig:return}b we show
the dependence of the total population of a community,
$B=\sum_{\ell=1}^L s_{\ell}p^{\ell}$, averaged over the recurrent set
$\mathcal{R}$, as a function of the resource saturation $R$ ($p^{\ell}$ denotes
the equilibrium density of the species at level $\ell$). The dependence is practically
linear, except for some dips near the onset of emergence of a new level.

\begin{figure}
\begin{center}
\includegraphics[width=80mm,clip=true]{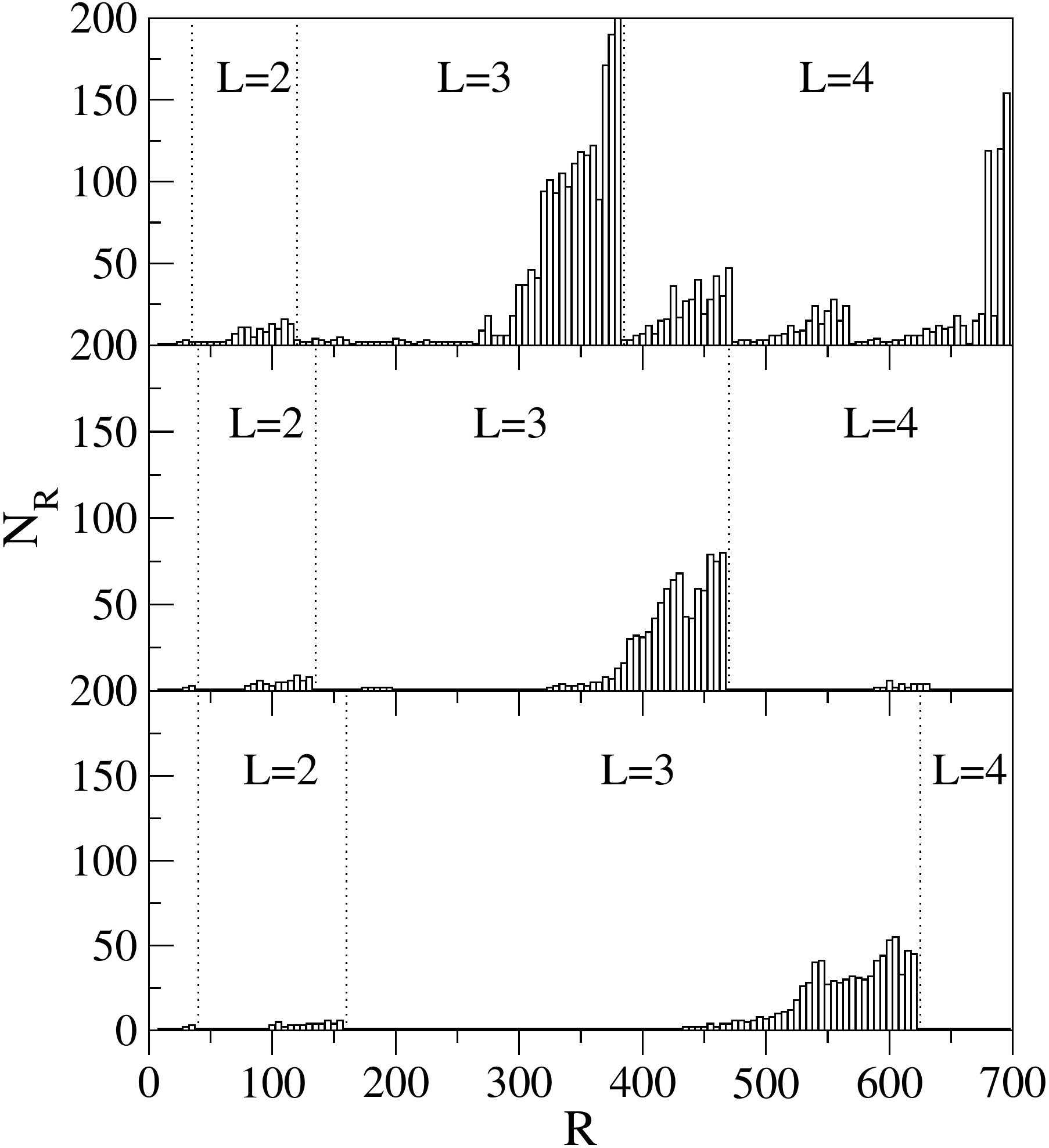}
\caption{Number of communities in the recurrent sets vs.\ the resource
saturation varying direct competition (upper panel, $\rho=0$; middle panel 
$=0.3$; lower panel, $\rho=0.7$). Observe the decrease in $N_{\mathcal{R}}$
as competition increases, and the increase of the values of $R$ at which
a new level arises.}
\label{fig:deprho}
\end{center}
\end{figure}

\subsection{Dependence on the parameters}
\label{ss:param}

We have already mentioned
that the model results are not qualitatively influenced by variations
of its parameters. For example, we have studied the model dependence 
with respect to direct competition (Figure~\ref{fig:deprho}). 
In the absence of interspecific competition ($\rho=0$)
levels are filled up more easily, so the number of
communities in the recurrent set increases with respect to the results
reported so far. The effect of increasing direct
competition is to reduce the number of ecosystems in these
sets, and to increase the resistance to the appearance of
a new level in the end state for the same
values of resource saturation. Thus, the global behavior of the number of
communities as a function of $R$ turns out to be similar, up to scale
factors, to that obtained in \cite{capitan:2009}, Figure~3.

\begin{figure*}
\begin{center}
\raisebox{44mm}{(a)}\includegraphics[width=82mm,clip=true]{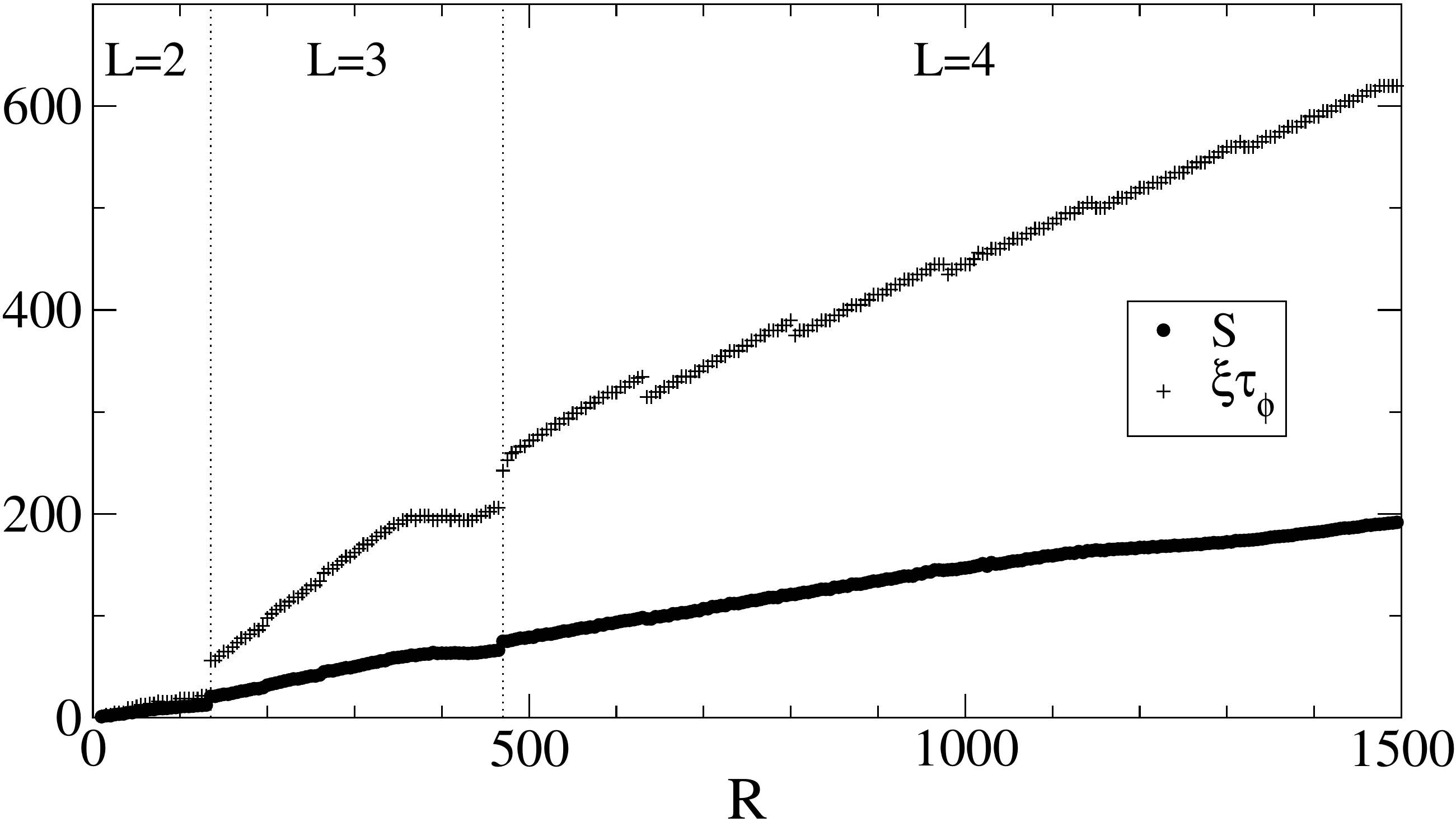}
\hspace*{4mm}\raisebox{44mm}{(b)}\includegraphics[width=85mm,clip=true]{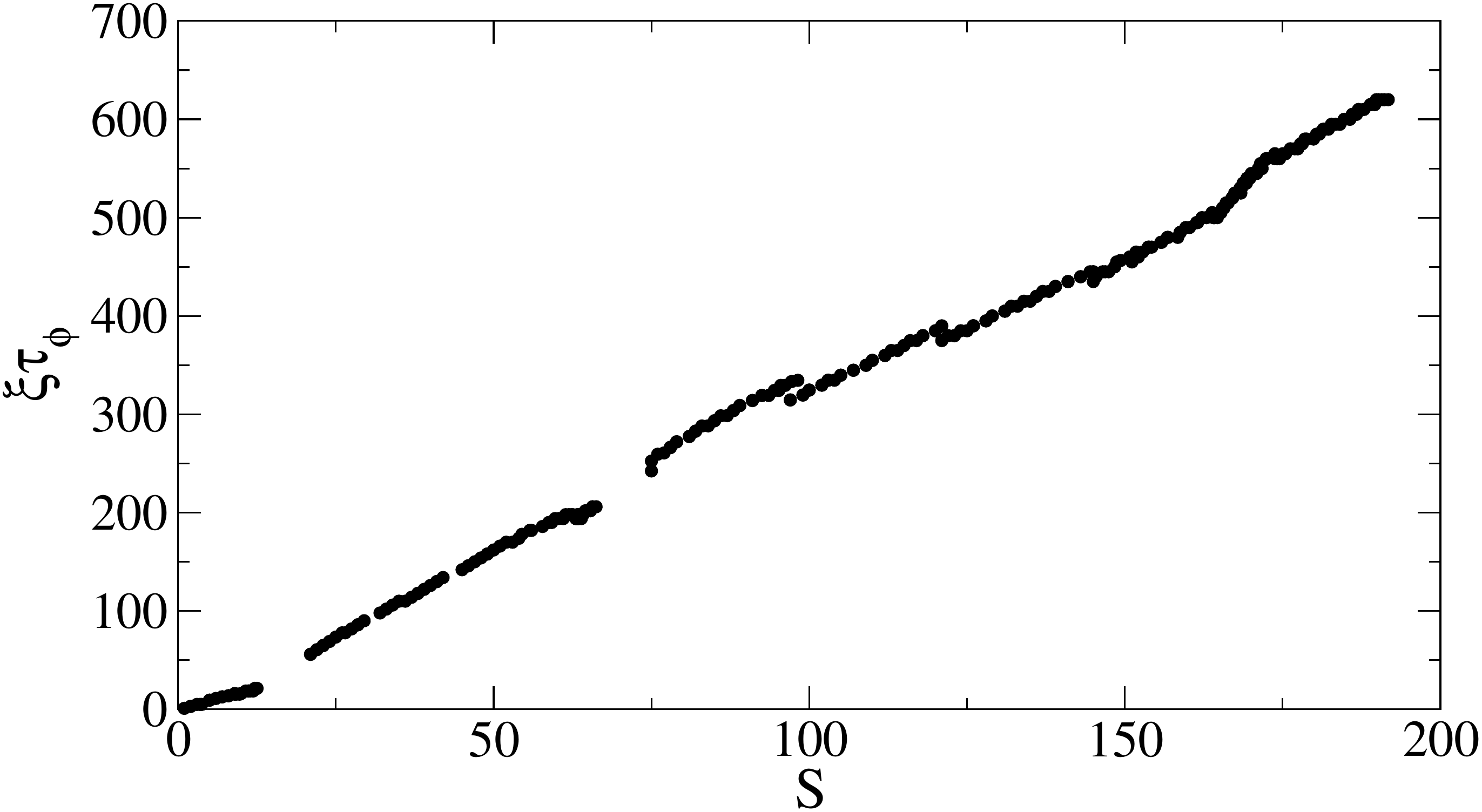}
\caption{(a) Mean absorption time $\tau_{\varnothing}$ (in number of invasions)
starting from the empty community $\varnothing$, and 
mean number of species $S$ as a function of the resource saturation $R$, showing a roughly
linear growth for both of them. (b) Mean absorption time $\tau_{\varnothing}$ vs.\ $S$.}
\label{fig:abstime}
\end{center}
\end{figure*}

The particular case $\rho=1$ (interspecific competition equal to intraspecific
competition) is qualitatively different. Fixing $\rho=1$ transforms the community into
a trophic chain. All species can be grouped into a single one with population
$N^{\ell}=s_{\ell} n^{\ell}$ [i.e. the Lotka-Volterra equations of this system 
\citep{capitan:2010a}
are closed in the variables $N^{\ell}$]. But the implications of this assumption
are stronger. Even if the distinction between species becomes meaningless,
one can formally keep the identities and treat them as different. But then it is easy
to show that any invasion attempted at a level already occupied by at least
one species will be unsuccessful because the population of the invader ends
up below $n_c$. In fact, according to Eq.~(12) of \cite{capitan:2010a},
the initial per capita growth rate of an invader at level $\ell$ is $-n_c$ and 
the equations for level ${\ell}$ and for the invader coincide. Hence 
$\dot{N^{\ell}}/N^{\ell}=\dot{n}/n$, $n$ being the abundance of the invader. This 
asymptotically yields
\begin{equation}\label{eq:equi}
\bar{p}=\frac{n_c}{N^{\ell}(0)}\bar{P}^{\ell}
\end{equation}
since $n(0)=n_c$ ($\bar{p}$ and $\bar{P}^{\ell}$ are the invader and the
total $\ell$-level densities at equilibrium after the
invasion, respectively). Now, the linear system (13) in \cite{capitan:2010a} for the interior 
equilibrium point before the invasion is exactly the same after the invasion
replacing $N^{\ell}(0)$ by $\bar{p}+\bar{P}^{\ell}$. This fact,
together with \eqref{eq:equi}, yields $p=n_c N^{\ell}(0)/[n_c+N^{\ell}(0)]<n_c$.
Since the population of the invader initially decreases, according to our extinction procedure 
\citep{capitan:2010a} the invader goes extinct.

Thus the assembly graph $\mathcal{G}$ becomes trivial. Using the notation
$\{s_{\ell}\}_{\ell=1}^L$ for each community, $\mathcal{G}$ is simply
\begin{equation}
\varnothing \rightarrow \{1,0,\dots,0\} \rightarrow \{1,1,\dots,0\} 
\rightarrow \dots \rightarrow \{1,1,\dots,1\}. 
\label{eq:chain}
\end{equation}
This never happens if $\rho\neq 1$. Things are thus very different when this fully
symmetric scenario is assumed. 

It can be shown that in this fully symmetric scenario the \emph{competitive
exclusion principle} applies. This principle states that there can not coexist more
populations than different resources (or ecological niches) 
in the long term if these populations depend linearly on the resources
\citep{hofbauer:1998}.
We can put this statement in mathematical terms.
For the sake of simplicity, let us assume that there is
a single trophic level with $S$ species predating on the resource 
(at rates $\gamma_{+i}$, $i=1,\dots,S$)
and let us set a non-uniform direct competition $\rho_{ij}$
between pairs of species in that level. Let $n_i$ be the population density
if species $i$, $a_i$ its death rate in the absence of consumption and 
$n_0$ the amount of resource. 
The Lotka-Volterra equations for this system are
\begin{equation}
\frac{\dot{n}_i}{n_i} = -a_i+\gamma_{+i}n_0-\sum_{j=1}^S \rho_{ij} n_j.
\end{equation}
If the competition matrix is singular, we can find a non-trivial solution 
$(c_1,\dots,c_S)$ for the linear system $\sum_ic_i\rho_{ij}=0$, $j=1,\dots,S$  
(note that, in particular, the fully symmetric scenario $\rho=1$ renders the
competition matrix singular). Hence 
\begin{equation}
\sum_{i=1}^S c_i(\log n_i)\dot{\phantom{i}} = \sum_{i=1}^S c_i (\gamma_{+i}n_0-a_i) 
\equiv -a
\end{equation}
where we can assume that $a$ is positive (otherwise change the sign of the
$c_i$). Integrating from 0 to $t$ we obtain
\begin{equation}
\prod_{i=1}^S n_i(t)^{c_i} = C e^{-a t}.
\end{equation}
This means that one of the densities must vanish in the limit $t\rightarrow \infty$,
which proves competitive exclusion.

There is a peculiarity of our model, though. If $\rho=1$ the population of
the invader at equilibrium will not be zero because in our model all constants are
uniform, so the equation to solve for $c_i$ is $\sum_i c_i=0$. This yields
$a=0$ and spoils the argument. However, we have shown that, with our procedure
of species extinction, the invader's population ends up below $n_c$ hence not
being viable. This restores competitive exclusion, albeit in a weaker sense.
The result \eqref{eq:chain} is just a manifestation of this fact.

It is important to notice that, for a non-singular competition matrix, the
competitive exclusion principle is not guaranteed to hold. In particular,
if $\rho<1$ the intra- and interspecific competition will have different magnitude,
and the matrix of elements $\rho_{ii}=1$ and $\rho_{ij}=\rho$ ($i\neq j$)
will be non-singular. The argument above does not apply anymore and, as a
matter of fact, by integrating the equations for population dynamics we
actually obtain more than one species coexisting with a single resource in the
system.

The interesting point brought about by the above discussion is that
interspecific competition induces \emph{de facto} a niche separation for
the species of the same level ---which are
therefore competing for the same resources--- that allows them to
circumvent the competitive exclusion principle [for a more thorough
discussion of this point see \cite{bastolla:2005a,bastolla:2005b}].

\subsection{Absorption times}
\label{ss:absor}

\begin{figure*}
\begin{center}
\raisebox{73mm}{(a)}\includegraphics[width=80mm,clip=true]{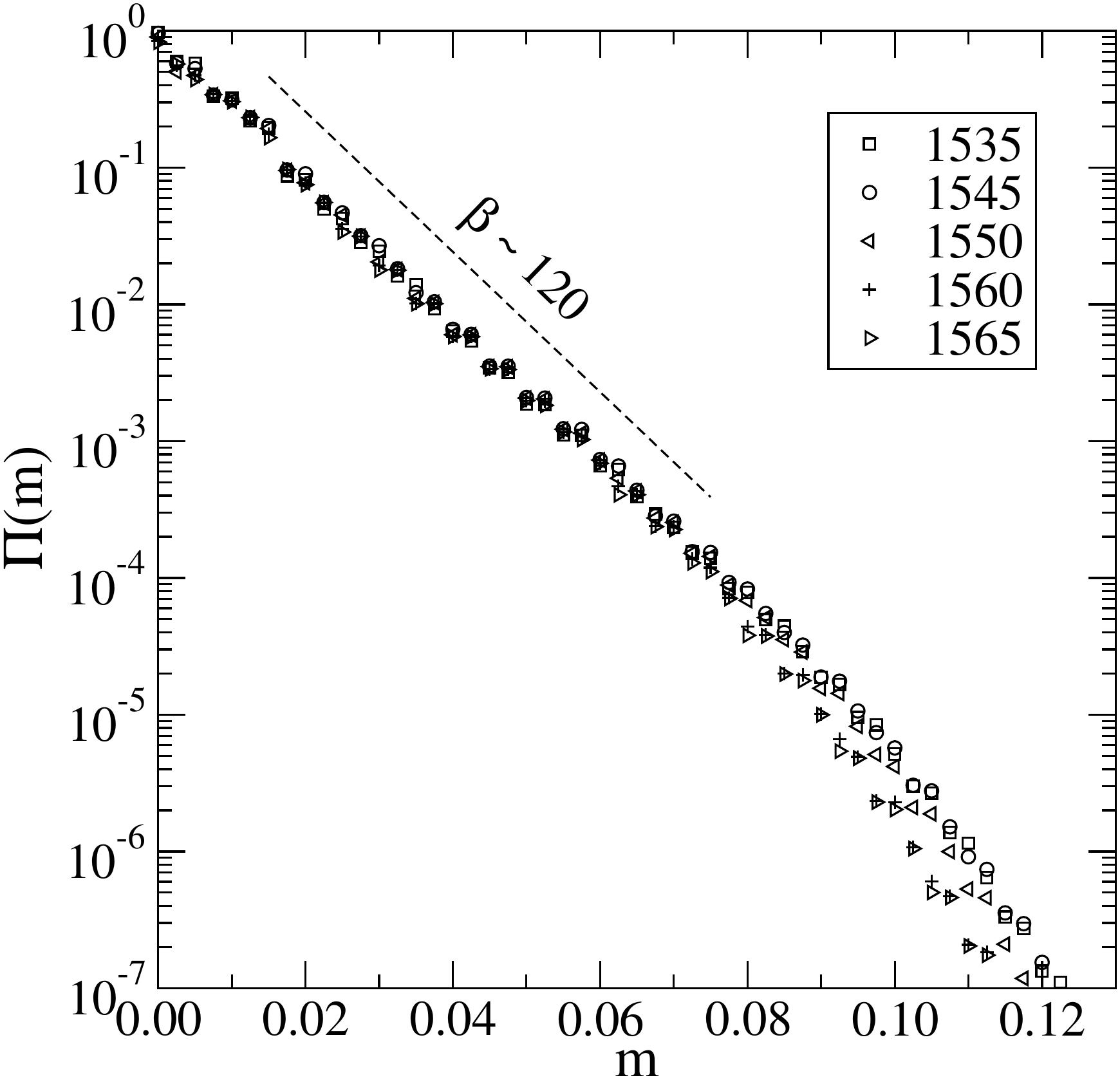}\hspace*{10mm}
\raisebox{73mm}{(b)}\includegraphics[width=84mm,clip=true]{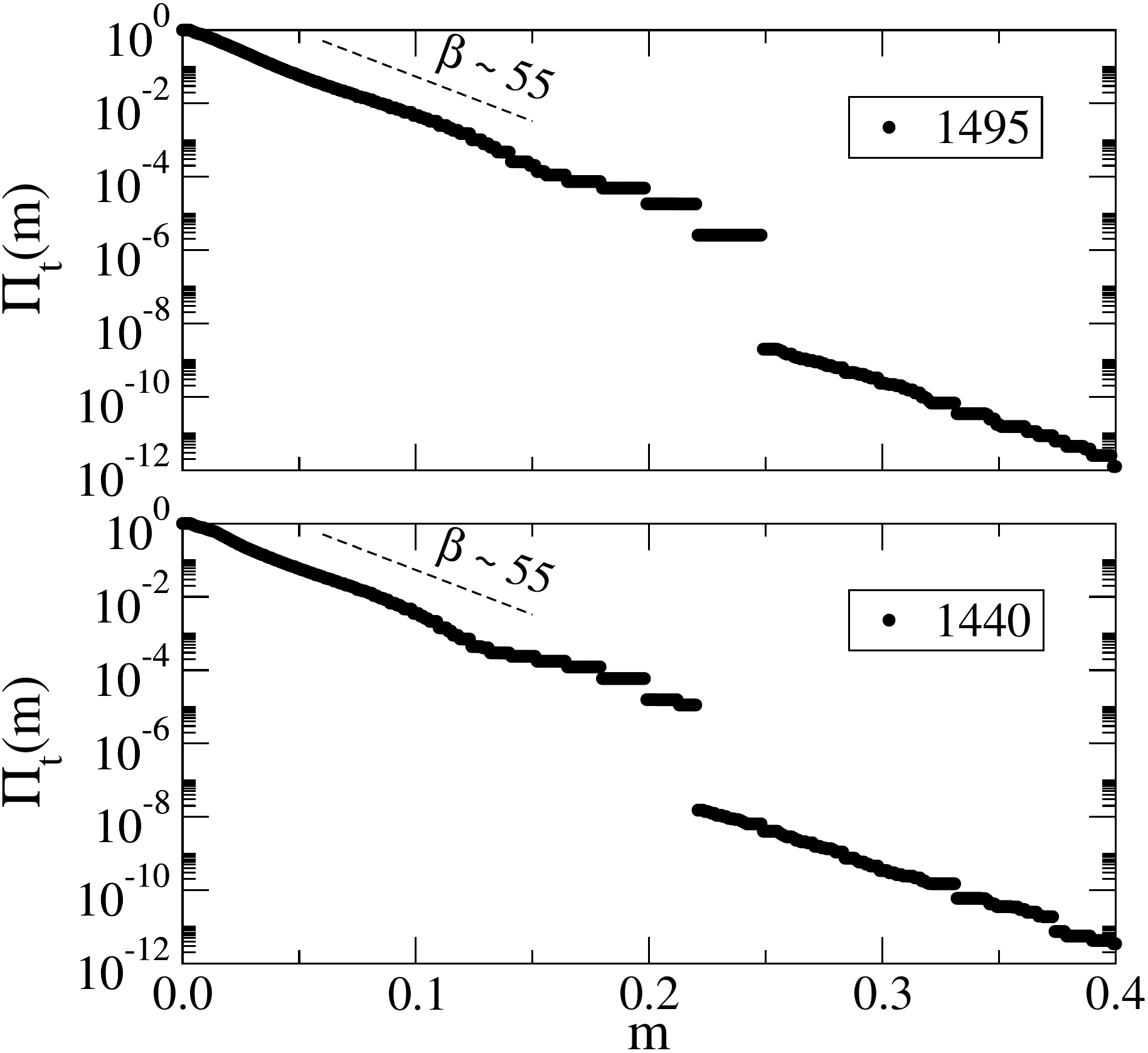}
\caption{(a) Probability $\Pi(m)$ that an invasion causes the extinction of at least
a fraction $m$ of the species of the invaded community, for several values of
the resource saturation $R$. There is a characteristic 
magnitude $\beta^{-1}$ around 1\%. 
In order to have enough statistics, we have chosen values of $R$
within the region where the number of communities in the recurrent set is above
1000 [see Figure~3 in \cite{capitan:2009}]. Thus we have
better statistics to compute the histograms than for smaller $R$.
(b) The same probability but for
transient states, $\Pi_{\rm t}(m)$, for two values of $R$. 
The characteristic size of the avalanches here is about 2\%.}
\label{fig:aval}
\end{center}
\end{figure*}
So far we have discussed properties of the recurrent set of the Markov chain 
associated to the assembly process. But we have not discussed the possibility
that the process may keep trapped for a long time in transient states. In order to ascertain
this, we have calculated the mean absorption time from the empty community 
$\varnothing$ to the end state. This can be done given the structure that
the transition matrix $P$ acquires after a permutation that reorders 
recurrent and transient states. We can thus write the matrix in a block form
\citep{karlin:1975}
\begin{equation}
P=\left(\begin{array}{cc}
U & 0\\
W & V
\end{array}\right),
\end{equation}
where matrix $U$ contains the transition probabilities within the recurrent set, and
$V$ contains transition probabilities between transient states. The average time
from the transient state $i$, to reaching 
for the first time the recurrent set at state $j$ is the element $t_{ij}$ of 
matrix $T$, where
\begin{equation}
T = \sum_{n=1}^{\infty} n V^{n-1} W=(\mathbb{I}-V)^{-2}W,
\label{eq:matrixT}
\end{equation}
$\mathbb{I}$ being the identity matrix. This expression counts as $n$ the
absorption time when the process remains $n-1$ time steps within the transient
subset and jumps to a recurrent state in the $n$-th step. The mean absorption time
for a process starting from the transient state $i$ will thus be
$\tau_i=\sum_{j\in \mathcal{R}} t_{ij}=(Tu)_i$, where $u=(1,\dots, 1)^T$.
Since $P$ is stochastic, $\sum_j (V_{ij}+W_{ij})=(Vu)_i+(Wu)_i=1$ for all 
$i \in \mathcal{G}-\mathcal{R}$, so $Wu=(\mathbb{I}-V)u$ or, equivalently,
$(\mathbb{I}-V)^{-1}Wu=u$. Together with \eqref{eq:matrixT} this implies
$(\mathbb{I}-V)\tau = u$, so solving this
sparse linear system yields the absorption times
for any transient state. Note that these times are proportional to $\xi^{-1}$,
because of the form \eqref{eq:mat} of our transition matrix.

In Figure~\ref{fig:abstime}a we plot the mean absorption time
$\tau_{\varnothing}$ to reach the recurrent set starting from 
the empty community, along with
the mean number of species $S$, which measures the size of
the system. Both of them grow almost linearly with $R$, hence $\tau_{\varnothing}$ is
roughly linear with $S$ as well (see Figure~\ref{fig:abstime}b). Since the 
number of states in each chain grows as $e^{\kappa \sqrt{R}}$, the number of states
of the Markov chain is very large compared to $\xi \tau_{\varnothing}$. 
Therefore the mean time to the end state is small compared to the system size.

This result should be taken with a grain of salt, because it strongly relies on
our assignment of probabilities to transitions. This, in turn,
assumes that there is always availability of invaders, which may not be true if
invaders come from a finite pool. The lack of potential invaders when the 
community is almost ``full'' would decrease the probability of a new invasion
and accordingly would increase the time that the process needs to reach the
end state. What the result of Figure~\ref{fig:abstime}a is actually telling us
is that the assembly graph is dominated by pathways in which most invasions are
accepted.

\subsection{Extinctions distribution}
\label{ss:extin}

As we have previously described, the assembly process can be regarded as
if the ecosystem self-organizes into a state resistant to invasions.
Either for transient or recurrent states, the
community is continuously undergoing avalanches of extinctions caused
by new colonizations. Figure~\ref{fig:aval}a shows a statistics of such
avalanches in some recurrent sets. It represents the probability $\Pi(m)$ 
that an invasion causes an
avalanche of magnitude greater than $m$ (understood as the fraction of
species that go extinct), averaged over the stationary state.
We can see in the figure that this probability shows an exponential decay,
with a typical avalanche size $\beta^{-1}$ of about $1\%$ of the community,
$\beta$ being the slope of the distributions in log-linear scale.
Invasions never cause big perturbations in the community.

\begin{figure*}
\begin{center}
\begin{minipage}[b]{0.5\textwidth}
\centering
\raisebox{36.5mm}{(a)}\includegraphics[width=88mm,clip=true]{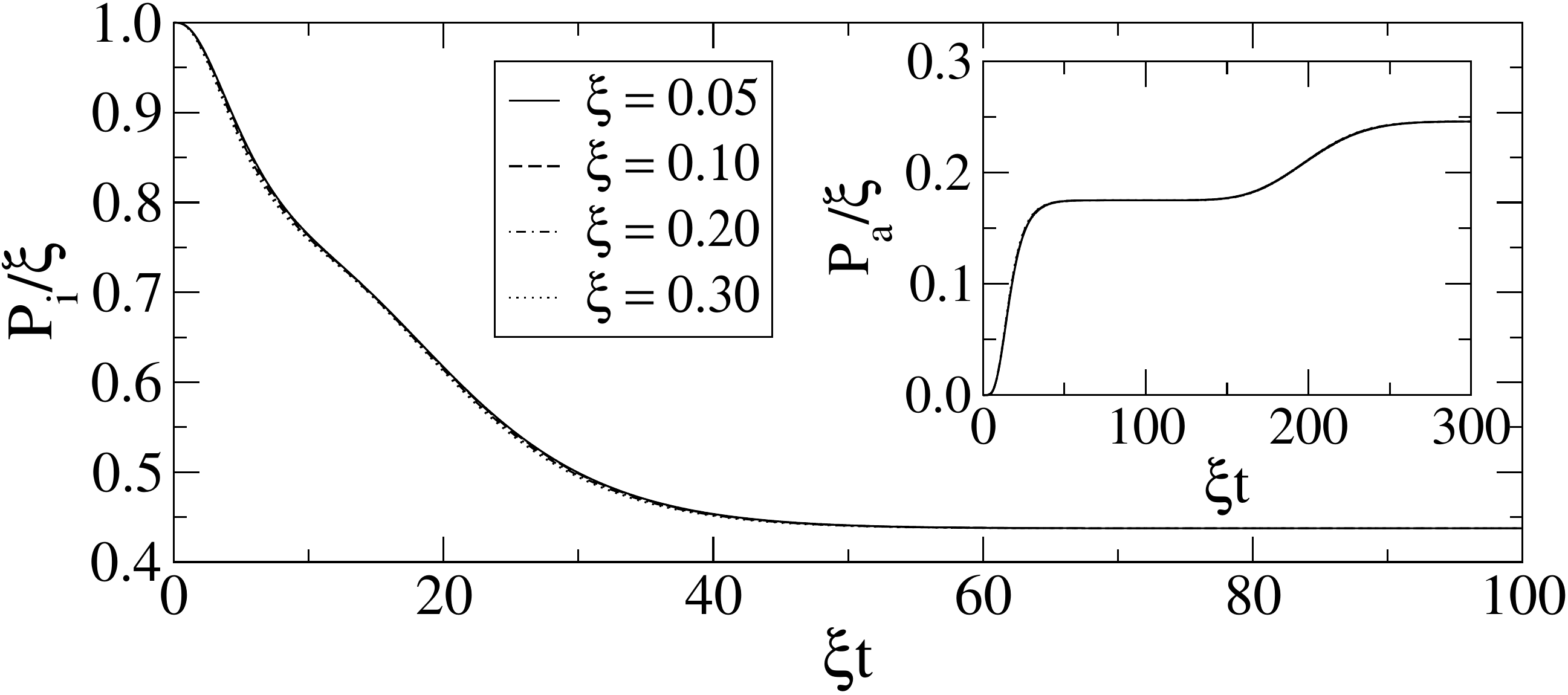}
\raisebox{36.5mm}{(b)}\includegraphics[width=88mm,clip=true]{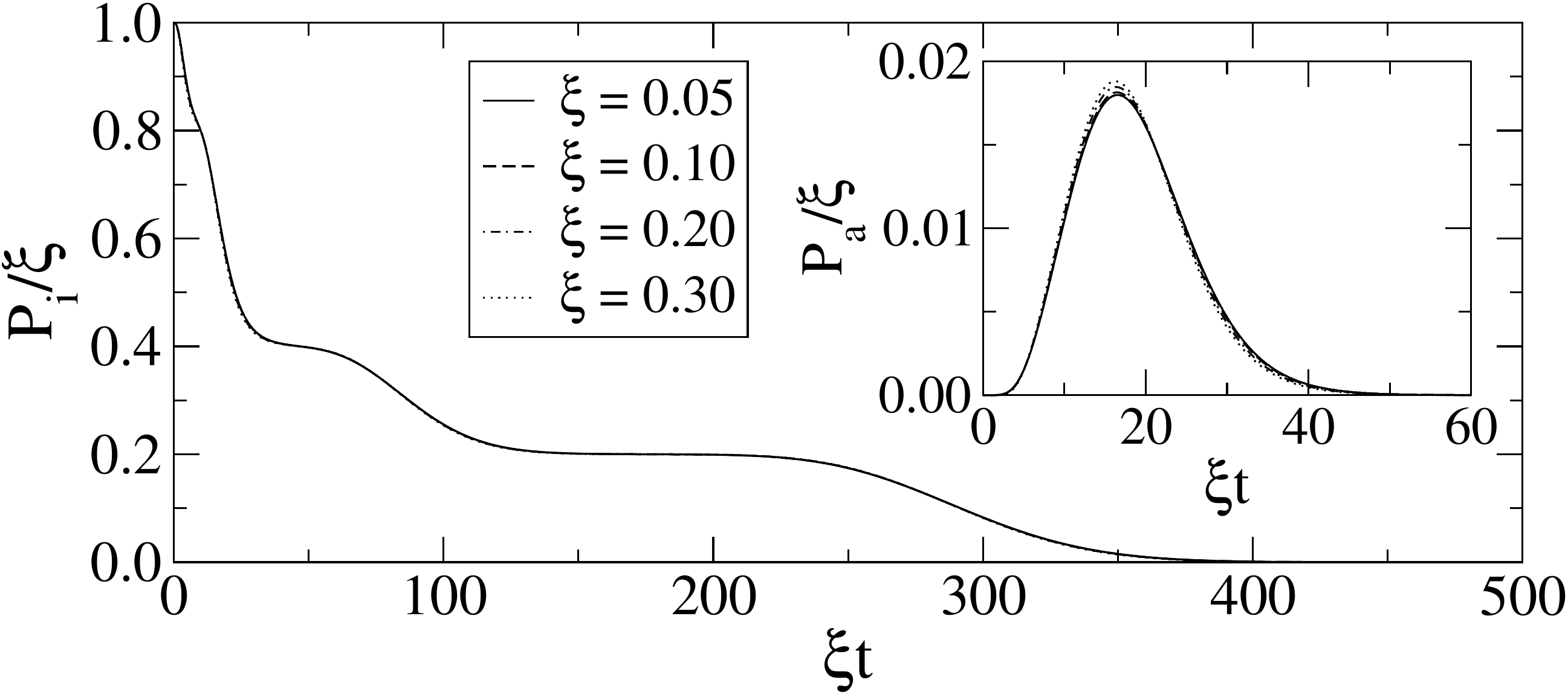}
\end{minipage}\hspace*{4mm}
\raisebox{76mm}{(c)}\includegraphics[width=83mm,clip=true]{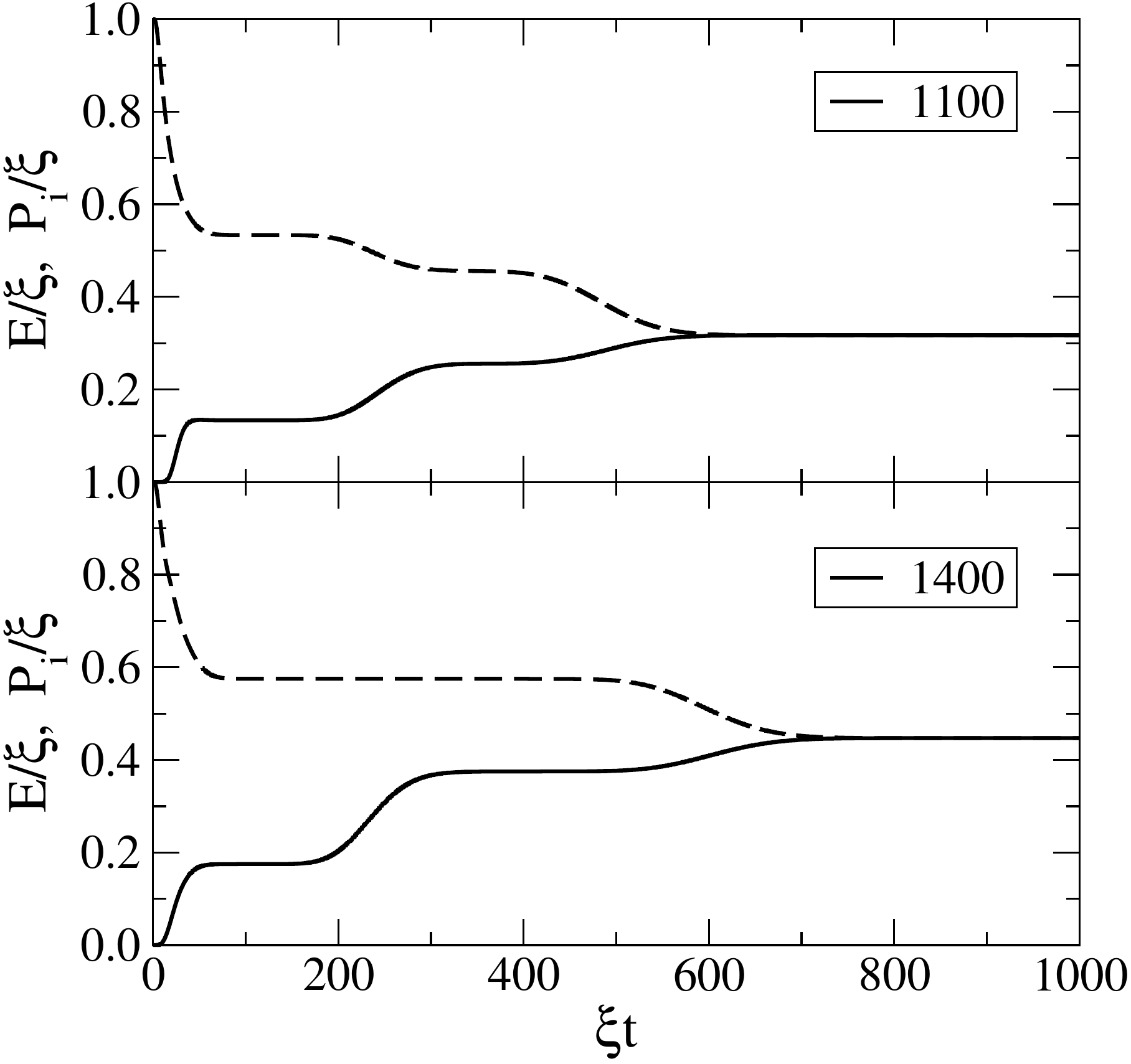}
\caption{(a) Probability of invasion vs.\ mean number of invasions ($\xi t$) 
for $R=430$ (with a complex end state).
Inset: probability of reconfiguration after invasion vs.\ mean number
of invasions.
(b) The same as (a) but for $R=540$ (the end state is a single community). (c)
Probability of invasion (dashed lines) and average number of species loss (full
lines) vs.\ mean number of invasions, for two values of $R$ with complex
end states. At the time these two magnitudes coincide,
a stationary state (the recurrent set) is reached.}
\label{fig:pinv}
\end{center}
\end{figure*}

We can calculate a similar distribution for the avalanches of extinctions in
the transient states. Now we have to weight the magnitudes with the average
fraction of visits to each transient state. Let us denote as $z_{ij}$ the average number 
of visits to state $j$ starting from state $i$. The matrix $Z=(z_{ij})$ is then
given by
\begin{equation}
Z = \sum_{n=0}^{\infty} V^n = (\mathbb{I}-V)^{-1}.
\end{equation}
Thus the number of visits to the transient $j$ starting from $\varnothing$ is
$\zeta_j=(u_{\varnothing} Z)_j$, $u_{\varnothing}$ being the row 
vector $\delta_{i\varnothing}$ (with a number
of entries equal to the number of transient states). We can calculate $\zeta$ by
solving the linear system 
\begin{equation}
\zeta (\mathbb{I}-V) = u_{\varnothing}.
\end{equation}

The resulting probability $\Pi_{\rm t}(m)$ that an invasion causes the extinction of at least
a fraction $m$ of the species of the invaded transient community is showed in
Figure~\ref{fig:aval}b. We also find an exponential behavior for the cumulative
distribution, in this case with a mean characteristic fraction of species loss of 2\%
for transient avalanches. The species loss caused by invasions in the transient
part of the graph is always small.

\subsection{Time averages}
\label{ss:ave}

Computing the time evolution of averages is very simple, given
the transition matrix $P$ and some initial probability distribution \citep{karlin:1975}
---which in our case is simply the vector $u_{\varnothing}$, since the
assembly process starts from the empty community. We just need to calculate
the power $P^t$ to obtain the transition probability matrix after
$t$ time steps.
Thus we can obtain the probability of rejecting the invader at discrete time $t$ as
\begin{equation}
P_{\rm r}(t) = \sum_j P_{jj} (P^t)_{j\varnothing},
\end{equation}
and that of accepting the invader as
\begin{equation}
P_{\rm i}(t) = \sum_j\bigg( \sum_{|k-j|=1} P_{jk}\bigg) (P^t)_{j\varnothing},
\end{equation}
where the inner sum runs over transitions from $j$ in which the invader is
accepted.
Obviously, the probability that the community undergoes a reconfiguration
because of the invasion is obtained as $P_{\rm a}(t)=1-P_{\rm r}(t)-P_{\rm i}(t)$.
Figures~\ref{fig:pinv}a and \ref{fig:pinv}b represent the dependence in time of
the probabilities $P_{\rm i}$ and $P_{\rm a}$
in two cases: one with a a complex end state (a), and another with a single 
community as end state (b). Notice that all curves collapse, for small $\xi$, when
divided by $\xi$ and plotted against $\xi t$ (mean number of invasions).

In Figure~\ref{fig:pinv}c we show the probability of invasion $P_{\rm i}(t)$ and
the average species loss defined as
\begin{equation}
E(t) = \sum_j\bigg( \sideset{}{'}\sum_k (\Delta S)_{jk} P_{jk}\bigg) (P^t)_{j\varnothing},
\end{equation}
where $(\Delta S)_{ij}$ is the species loss in the transition from $j$
to $k$ and the prime denotes that we ignore in the sum transitions
in which the invader is accepted. When these two magnitudes are equal
there is an equilibrium between the average frequency of invasions and the
average number of species loss. This is a fingerprint of the reaching of
the stationary state. As expected, this time is comparable to the absorption
time shown in Figure~\ref{fig:abstime}a.

\begin{figure*}
\begin{center}
\raisebox{46mm}{(a)}\includegraphics[width=84mm,clip=true]{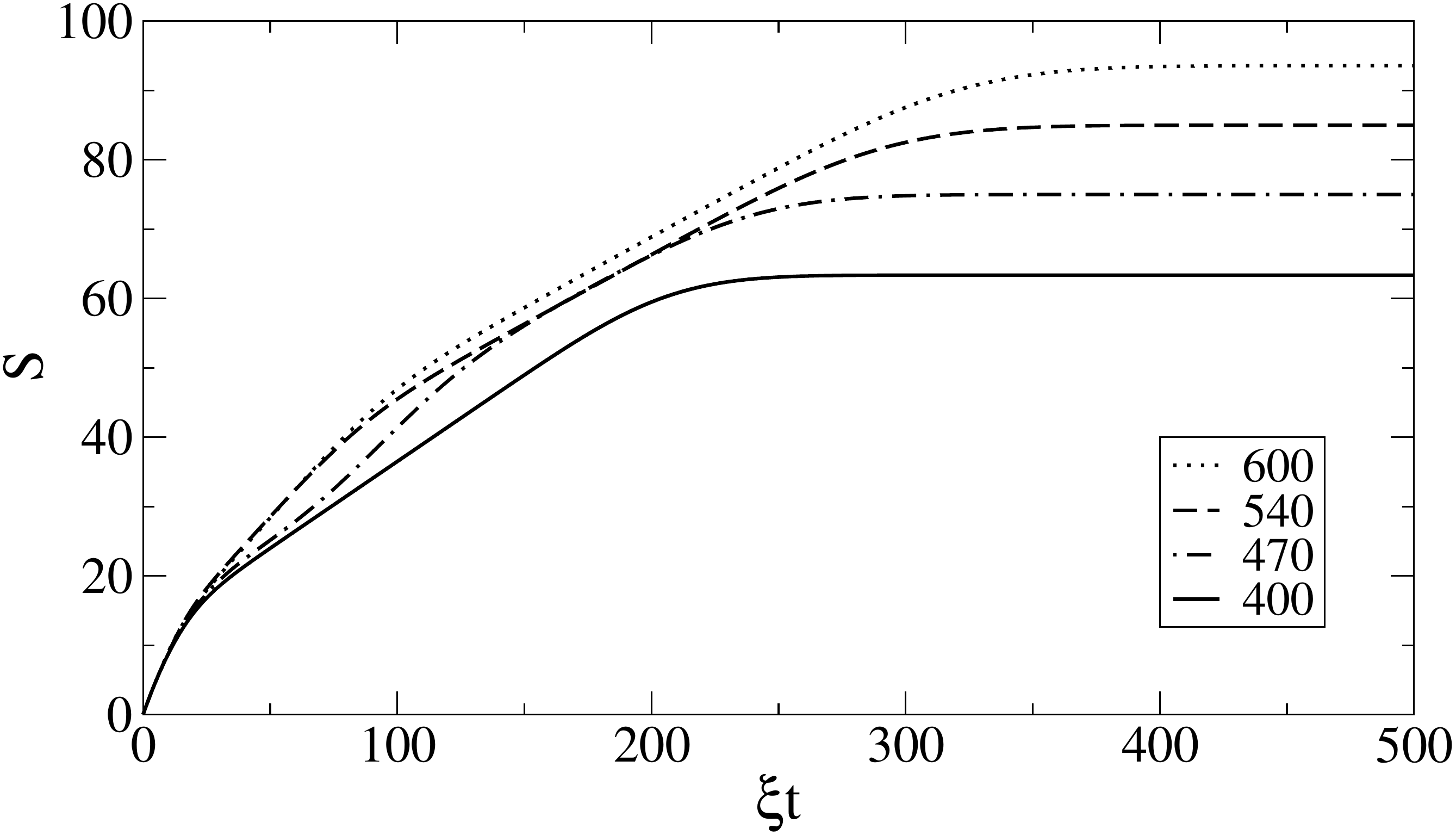}
\hspace*{4mm}\raisebox{46mm}{(b)}\includegraphics[width=84mm,clip=true]{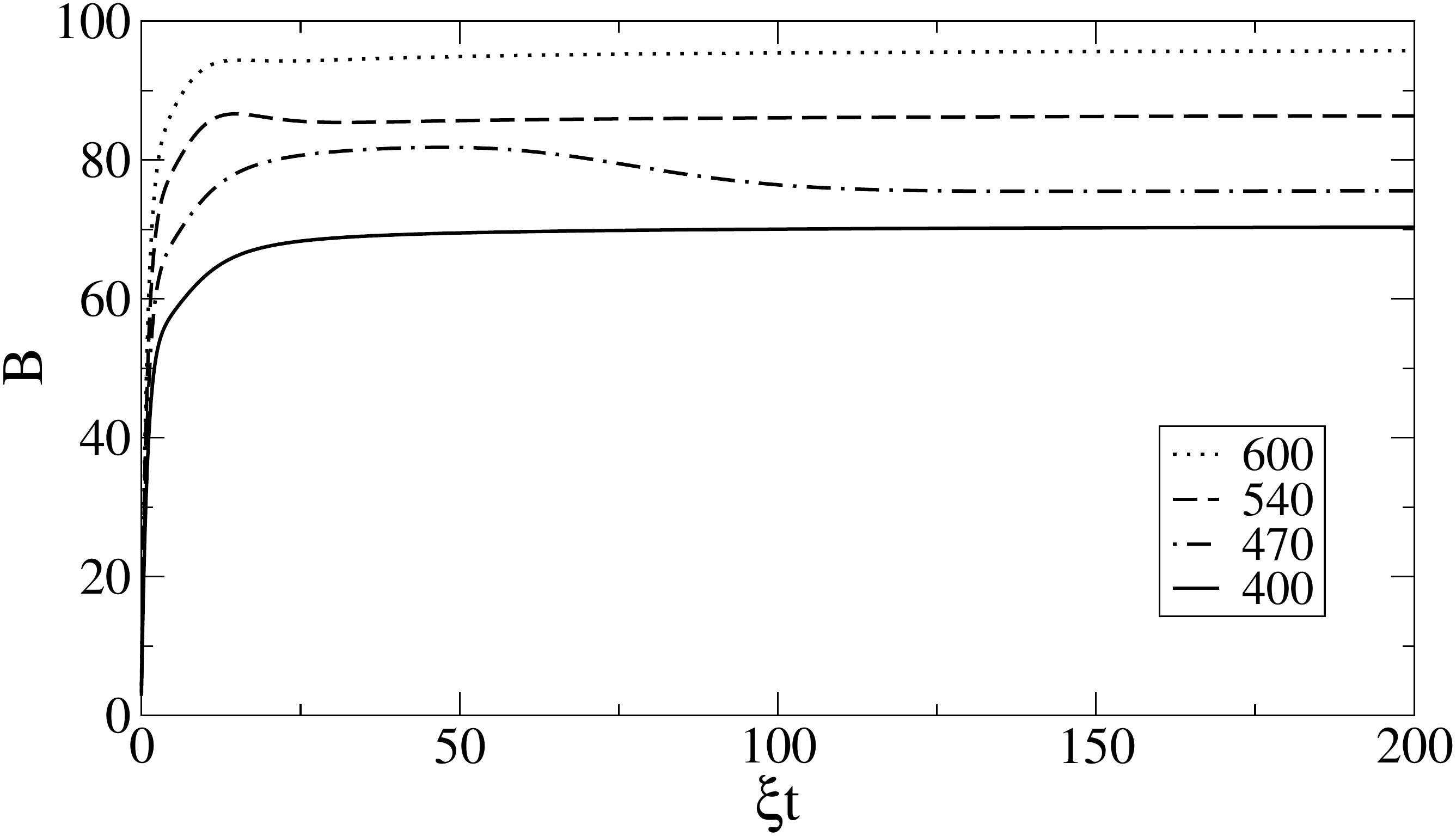}
\raisebox{45mm}{(c)}\hspace*{3mm}\includegraphics[width=79mm,clip=true]{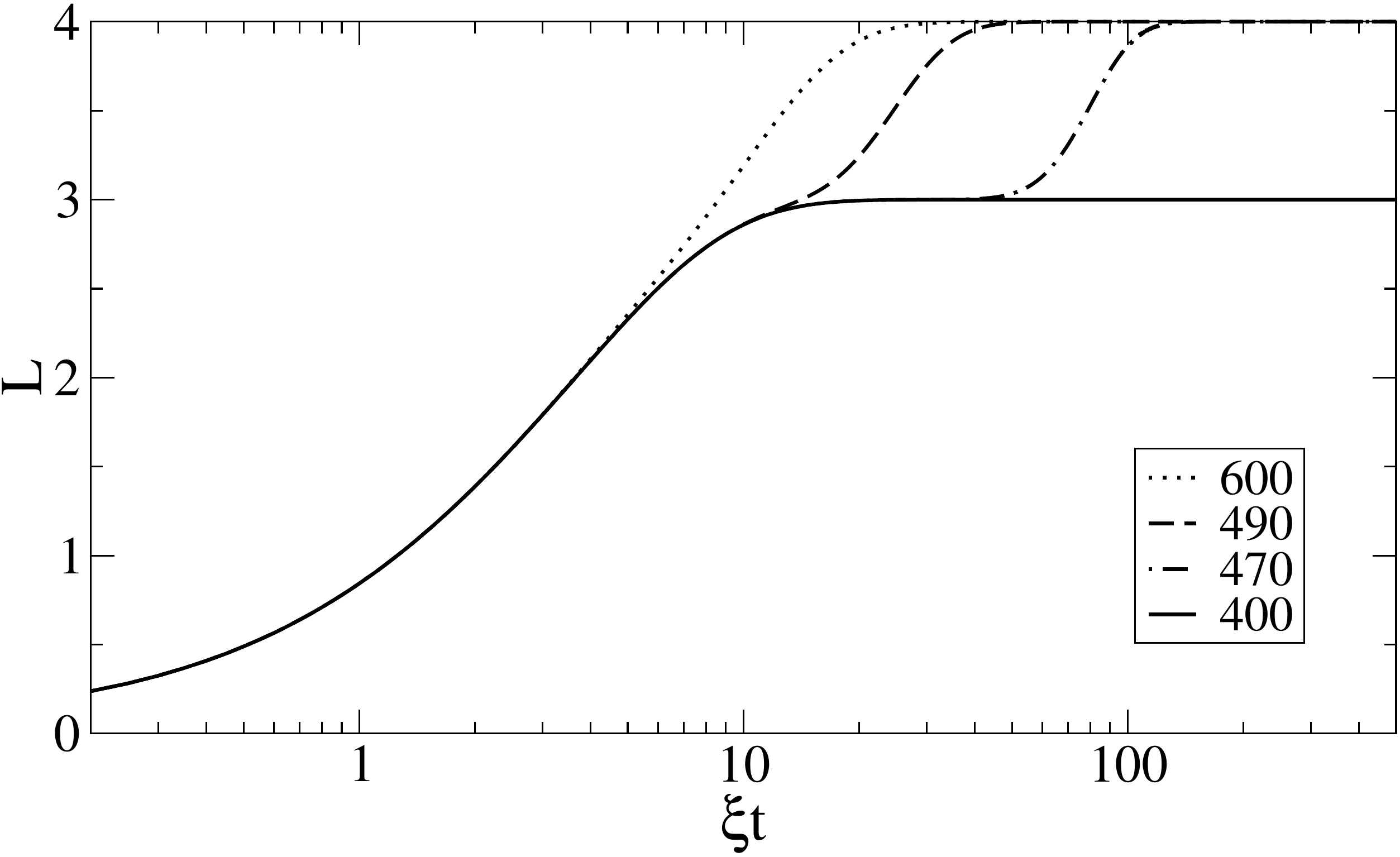}
\hspace*{6mm}\raisebox{45mm}{(d)}\includegraphics[width=84mm,clip=true]{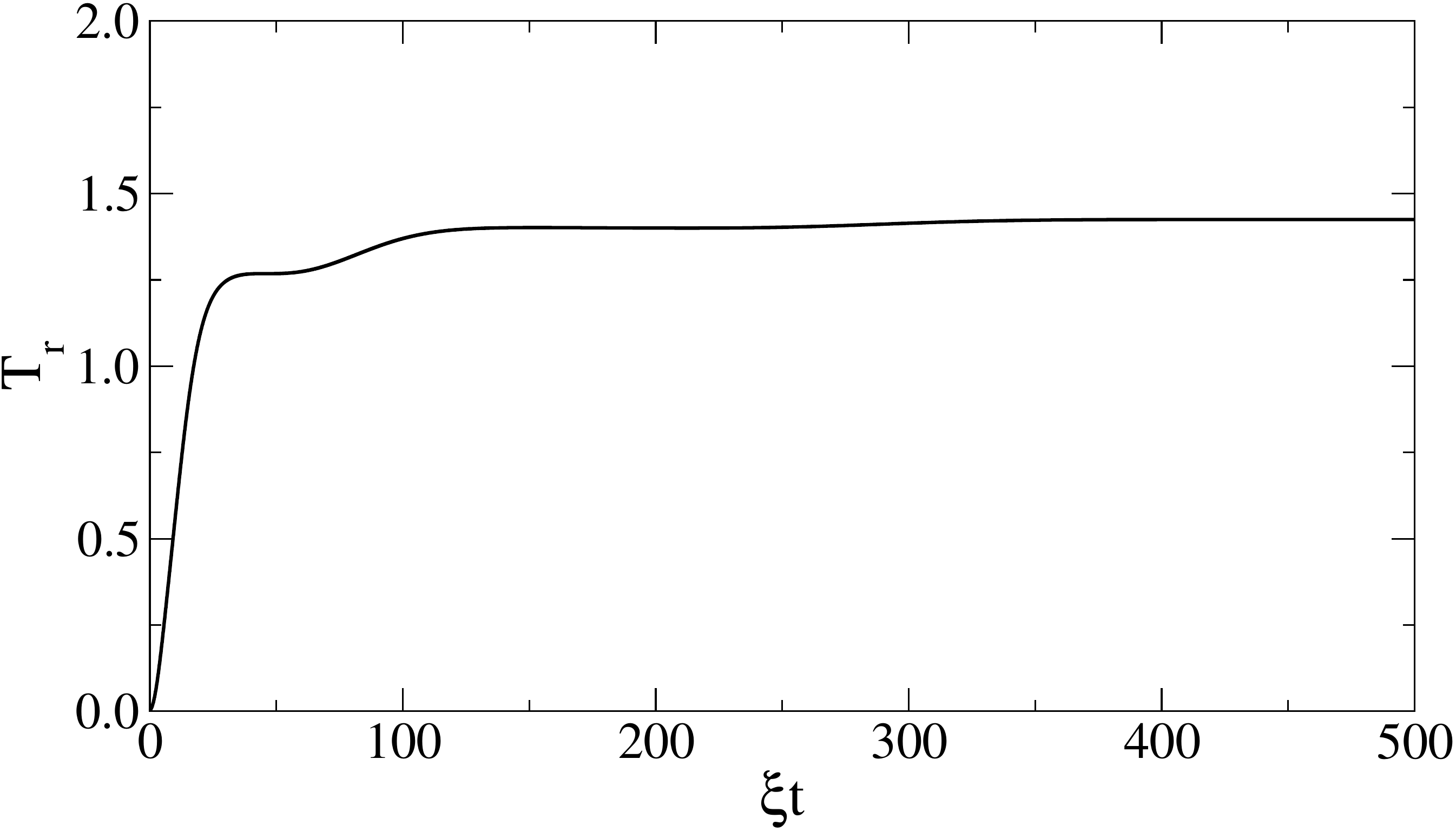}
\caption{Species richness (a) and total population (b) vs.\ mean number of invasions
($\xi t$), for several values of the resource saturation $R$.
At $R=470$ (showed with dash dotted lines) 
the ecosystem crosses over from 
$3$ levels to $4$ levels (this crossover corresponds to the non-monotonic
behavior of the total population).
(c) Mean number of levels vs.\ mean number of invasions. At the onset
of the fourth level, the ecosystem stays some time trapped in three-level communities.
(d) Typical variation of the average return time with $\xi t$.}
\label{fig:bio_time}
\end{center}
\end{figure*}

Another important magnitude is biodiversity. Figure~\ref{fig:bio_time}a represents
the evolution of the average number of species for several values of $R$. 
In all cases, this average number grows monotonically
until reaching the stationary state, so biodiversity and resistance to invasion are
positively correlated, in agreement with previous assembly models
\citep{law:1996,morton:1997}.

Figure~\ref{fig:bio_time}b
represents the evolution of the total population density $B$ of each community.
If we assume, for the sake of simplicity, the same weight
per individual for all species in our model communities, then $B$ can be regarded
as the total biomass in the community. 
Although there is a clear trend for biomass to increase, it is not always
at its optimum in the stationary state. This is very clear in the figure for
$R=470$, a value at the onset of the appearance of the fourth trophic level. This agrees
with the analysis performed by \cite{virgo:2006} on their assembly model.

We have also studied the time dependence of the average number of trophic levels during
the assembly, which is shown in Figure~\ref{fig:bio_time}c. At $R=470$ the process
stays a certain time trapped in three-level communities until the fourth level
is finally accepted. This effect becomes lower upon increasing $R$, until
there is no trapping and the fourth level is reached straight away.

Figure~\ref{fig:bio_time}d shows a typical time evolution of the average return
time along the assembly until reaching the stationary state. Communities are
less resilient (have larger return time to equilibrium) as time increases. Thus,
there is a trade-off between robustness (resistance against invasions) of the
ecosystem and dynamic stability which is resolved by sacrificing the latter
in favor of the former.

\section{Discussion}
\label{s:discussion}

In this work we have provided a full account of results of the model introduced in
\cite{capitan:2009}. The results presented here have been obtained from a direct
analysis of the Markov chain describing the assembly process. Among the novel results 
here presented are the dependence of many biological observables on the amount of available
resource (implemented through parameter $R$), the average times that the process
needs to reach the recurrent set, the statistics of avalanches in both transient
and recurrent states, and the time evolution of any observable.

Our model might be considered as a benchmark of the assembly 
process that builds up ecological communities. As such, we do not aim at providing a
realistic description of an ecosystem but at capturing, in a very simplified model,
the essential mechanisms that do occur in the construction of real ecosystems.
The model rests on some oversimplistic features: communities are strictly organized
in levels, predation occurs only between contiguous levels, 
species of a given level are trophically equivalent, model parameters
are chosen uniformly and the population dynamics is ruled by simple Lotka-Volterra
equations. In spite of this, our model exhibits the same behavior as all other
assembly models reported in the literature. This indicates that this behavior
is very robust, and probably shared by real systems and simple models alike. 

Thanks to these oversimplifications the model provides important advantages
on previous assembly models. The main one is that we can trace all pathways
of the assembly process. This allows us to compute exactly all the observables
of a community and to characterize in a very precise manner the stationary
state of the ecosystem. Our model also has a species pool, as standard assembly
models, but because we allow for
an arbitrary number of trophically equivalent species, the pool is infinite
and the model does not suffer from the problem of exhaustion of good invaders
that may trap the community in a transient state \citep{case:1991,levine:1999}.
This has permitted us to build communities with hundreds of species and explore
the influence of different elements on the behavior of the assembly process.  

Therefore, we are not limited, as in standard assembly
models, to compute averages over a set of realizations of the process. 
As we pointed out in \cite{capitan:2009}, the number of shortest pathways
leading from $\varnothing$ to the recurrent set can be enormous. For instance,
for $R=300$ (a case with an absorbing community of three trophic levels and 50
species), there are $\sim10^{10}$ different minimum-length pathways. This
number is far from anything a simulation can come close to.

There is, of course, a concern about having trophically equivalent ---hence
indistinguishable--- species. The grouping of trophically equivalent species
is a common practice in studying food webs, so it is tempting to do so in
this model. If we do it, the model becomes equivalent to a chain, for which
Lotka-Volterra dynamics is well characterized \citep{hofbauer:1998}, and the
invasion process seems to become trivial. This is not true, though:
if $\rho\ne 1$, i.e. if intra- and interspecific competition are different in magnitude, 
intraspecific competition in the
equivalent chain explicitly depends on $s_{\ell}$, so invasions modify
the parameters of the chain and the invasion process becomes non trivial.
Thus, it is because of the direct interspecific competition $\rho<1$ that this
equivalence breaks down and the model departs from triviality.
We have explicitly shown that choosing $\rho=1$ brings about the
competitive exclusion principle, and indeed the model turns into a chain.
But for any $\rho<1$ this does not longer hold. Interspecific competition is
thus an effective way of creating new niches.

Let us now summarize the main conclusions we can extract from the present
analysis of the model.

As our model ecosystems evolve we observe three trends: biodiversity increases, resistance
to invasion increases and all species decrease their populations. In the steady
state biodiversity is at its maximum, all populations are close to the extinction
level and either invasions are rejected or they produce transitions between a
set of communities with a very similar structure. All three features are related.
The increase in biodiversity is unavoidable because of the constant flux of
colonizers; however, as the number of species increases, their populations 
necessarily decrease because all share the same resource. The invasion process
guarantees that this is done in the most efficient way, because inefficient
invasions cause extinctions in the community and force a more equilibrated
rearrangement of the populations. This, in turn, justifies the increasing
resistance to new invasions. At the end, all populations are so close to
extinction that either no new invasions are possible, or they just cause
small rearrangements that leave the community in a similar state.

Final communities have typically three or four trophic levels ---only 
ecosystems with more than $200$ species generate
five trophic levels. On the other hand, the number of species in each level
has a pyramidal structure. Both features are in qualitative agreement with
what is observed in real ecosystems \citep{cohen:1990} and we have
discussed at length the properties of the population dynamics equations that 
explain these features in \cite{capitan:2010a}.

As already advanced in \cite{capitan:2009} the end state is always
unique, and this is consistent with previous assembly models \citep{morton:1997}.
However, there is a caveat that should be made on this point related
to the indistinguishability of species within the same trophic level:
the end state is unique as long as we consider only the number of species at
each level. Whether two communities with the same numbers have the same
or different species is meaningless for this model, so the conclusion
is not definitive. In fact, some relatively recent experiments on
aquatic microbial communities 
establish that productivity-biodiversity relationships depend on the history of assembly
\citep{fukami:2003}, and it is our guess that the independence on history
resulting from this model might be an artifact of the indistinguishability of
species. Refined versions of this model may clarify this issue.

As for the robustness of the above results, we have tried other values of the
direct competition parameter, namely $\rho=0$ and $\rho=0.7$, to test
its influence. No qualitative difference with the behavior reported here
is found. Nonetheless, there are three quantitative effects that we have
observed
as $\rho$ increases: resistance to invasion increases, appearance of new
trophic levels is hindered and the number of communities in complex
end states decreases. Varying $\gamma_-$ has similar effects; in fact,
the product $\gamma_+\gamma_-=0.1\gamma_-^2$ provides a quantitative
estimate of indirect competition.

It can be argued that parameters should depend on the trophic level rather
than being uniform for all species. It is very easy to show that this does
not change the dynamic stability patterns because in that case one can also
construct a Lyapunov function [see \cite{capitan:2010a}]. We have
not attempted any test in this respect, but it is hard to believe that such
a variant of the model will produce any qualitative difference.
The assembly graphs will be similar to the
ones found for the present model. Something more can be said about the invasion
rate. We have presently assumed that the invasion probability is the
same for all trophic levels, but notice that the assembly graph is
utterly independent on this choice, so certainly choosing a different
invasion probability will change the numerical value of the nonzero
entries of the transition matrix $P$, but only them.
The graph, as well as the structure of transient and
recurrent states of a finite Markov chain, only depends on which
elements of $P$ are zero \citep{karlin:1975}, so not just the graph
but the set of communities in the end state will be exactly the same
as those reported here (the probability distribution in the steady state
will, of course, be different).

Perhaps the most important limitation of this model is the choice
of the Lotka-Volterra equations. The choice of population dynamics has been
reported to have a strong influence in the final shape of ecological communities
\citep{drossel:2004,lewis:2007}. Introducing non-linear equations leads to
more complex stability patterns than simply rest points. How to account for
them is not yet clear to us, but neither is whether this will really affect
the qualitative behavior of the assembly process. Thus, this remains an
important open question that deserves further analysis.

There are further open questions such as the application of this model to 
metacommunities. The resulting ecosystems
can be readily altered when migration takes place among spatially
distributed patches. With a simple model like
ours, it might be possible to build up an assembly graph between different
communities in different patches. The interplay between communities 
in different patches could lead to an outcome different from the one we
obtain with a single patch. On the other hand, a simple model like this can
provide us with basic understanding of complex processes such as, for instance,
the rebuilding of a natural community after its degradation. Very little is known
about the processes that helps to reconstruct damaged communities, and a
simple framework like ours could provide some hints about how to tackle this
problem from a theoretical point of view.

The final take-home message from this work is this: we should not be
afraid of oversimplifications in complex systems. Complexity normally
arises as a consequence of a collective behavior of many entities, not
as a result of the complexity of interactions. The key point is whether
we are retaining the basic ingredients yielding the desired output. We
have shown that there is no qualitative difference between the results
of this oversimplified model and previous, more sophisticated assembly
models. And there is a lot to gain from the wider view that this model
provides of the process and the much higher control we have on the
parameters. Many questions that are hard (or even impossible) to answer
in previous model have a clear-cut answer here. And even if they may
be too simplistic, they can still guide our intuition when dealing
with real ecosystems.

\section{Acknowledgements}
This work is funded by projects MOSAICO, from Ministerio
de Educaci\'on y Ciencia (Spain) and MODELICO-CM, from Comunidad Aut\'onoma de Madrid
(Spain). J.\ A.\ Capit\'an also acknowledges financial support through a contract
from Consejer\'{\i}a de Educaci\'on of Comunidad de Madrid and Fondo Social 
Europeo. J.\ B. is funded by the European Heads of Research Councils, the
European Science Foundation, and the EC Sixth Framework Programme through a EURYI
(European Young Investigator) Award.

\appendix

\bibliographystyle{model2-names}
\bibliography{ecology}







\end{document}